\documentclass[12pt,pra,reprint,superscriptaddress,amsmath,amsfonts,amssymb,nofootinbib]{revtex4-1}

\usepackage[utf8]{inputenc}
\usepackage[T1]{fontenc}
\usepackage[sc,osf]{mathpazo}
\usepackage{amsmath, amsthm, amssymb,amsfonts,mathbbol,amstext}
\usepackage{graphicx}
\usepackage{dcolumn}
\usepackage{bm}
\usepackage{bbm}
\usepackage{mathtools}
\usepackage{comment}
\usepackage{multirow}
\usepackage{diagbox}
\usepackage{float}

\usepackage{braket}
\usepackage[english]{babel}
\usepackage{graphicx}
\usepackage[caption=false]{subfig}
\usepackage{mathrsfs}
\usepackage{amsmath}
\usepackage{amsthm}
\usepackage{dsfont}
\usepackage{thmtools, thm-restate}
\usepackage{bm}
\usepackage{xcolor}
\usepackage{hyperref}
\usepackage{hypernat}
\usepackage[utf8]{inputenc}
\usepackage[T1]{fontenc}
\usepackage{cleveref}
\usepackage{soul}
\usepackage[normalem]{ulem}

\theoremstyle{plain}
\newtheorem{theorem}{Theorem}[section]

\theoremstyle{definition}
\newtheorem{definition}[theorem]{Definition}

\theoremstyle{remark}

\renewcommand\bra[1]{{\langle{#1}|}}
\renewcommand\ket[1]{{|{#1}\rangle}}
\newcommand{\ketbra}[2]{\ket{#1}\bra{#2}}

\renewcommand{\braket}[2]{\langle #1 | #2 \rangle}

\begin{document}
\title{Agreement and Compatibility in Wigner's Friend Paradox}

\author{Julio C. F. Silva}
\email{julio.fernandes@estudante.ufjf.br}
\affiliation{Depto. de F\'isica, ICE, Universidade Federal de Juiz de Fora, MG, Brazil}

\author{B. F. Rizzuti}
\email{brunorizzuti@ufjf.br}
\affiliation{Depto. de F\'isica, ICE, Universidade Federal de Juiz de Fora, MG, Brazil}

\author{Cristhiano Duarte}
\email{cristhianoduarte@gmail.com}
\affiliation{Depto. de F\'isica, ICE, Universidade Federal de Juiz de Fora, MG, Brazil}
\affiliation{Instituto de Física, Universidade Federal da Bahia, Campus de Ondina, Rua Barão do Geremoabo, s.n., Ondina, Salvador, BA 40210-340, Brazil}
\affiliation{Fundação Maurício Grabois, R. Rego Freitas, 192 - República, São Paulo - SP, 01220-010, Brazil}
\affiliation{Institute for Quantum Studies, Chapman University, One 
University Drive, Orange, CA, 92866, USA}

\begin{abstract}

There has been an upsurge of interest in the consequences for quantum physics of the so-called Wigner's Friend Paradox. In its original formulation, the paradox has been turned inside out, and virtually every aspect of it has been looked into. Consequently, it is becoming widely accepted that we can find the potentially puzzling consequences of Wigner's thought experiment only in light of its many-parties extensions. Nonetheless, this contribution returns to the source. Reframing the question as an inference problem, we advance a radically Bayesian interpretation that shows no contradiction between Wigner's and Wigner's Friend's descriptions---neither classically nor quantumly. Therefore, with no paradoxical conclusion. In doing so, we flesh out and expose previously untouched aspects of Wigner's thought experiment, in particular, the fact that compatibility and agreement are fundamental to our understanding of it, and that a rigorous notion of compatibility is the key to paving the way for the agents to reach an agreement. Also, by conservatively extending Wigner's original setup and incorporating what we call the 'benefit of the doubt', we see how either Wigner's or his Friend's description can be driven to match one another's---an impossibility if either agent does not keep an open mind.

\end{abstract}

\maketitle

\section{Introduction}\label{Sec.Intro}

Quantum theory comprises a set of rules intended to deal with actual experiments whose alternative classical descriptions fail to work as expected~\cite{JL16}. On a more practical dimension, quantum physics has several unique features that can be exploited in the run to new and disruptive technologies~\cite{Preskill18}: no-go theorems like the no-cloning and no-broadcasting theorems have been translated into essential ingredients for the security of quantum cryptographic protocols~\cite{HJP23}; puzzling aspects of the mathematics of quantum theory, like entanglement and superposition, are now part of the canon of standard quantum information and quantum computation tasks~\cite{NC10, Horodecki09}; and, more recently, we have also seen subtler aspects derived from the machinery of quantum theory as resources to fuel tasks with either no classical counterpart or with substantial gains in efficiency~\cite{CG19, BMPM23, MottaEtAl20, AruteEtAl19}. What is, in general, less advertised is that, notwithstanding these real-world implications, there remain open questions deeply seated in its foundations.

Consider, for example, the thought experiment designed by Eugene Wigner in 1961~\cite{Wigner95, FreireJunior2015}. According to him, there is an apparent paradoxical incompatibility in the descriptions of either agent present in the scenario~\cite{Cavalcanti21}---we will address the \emph{Gedankenexperiment} and its apparent contradiction in more detail in a minute. What is important is that even though Wigner's original thought experiment was conceived within the debate around the mind-and-body problem, the question it advances is closely connected to the measurement problem~\cite{Schlosshauer05} and exemplifies some fault lines still open in the foundations of quantum mechanics. Other foundational problems include the status of the wave function~\cite{Leifer14}; whether it is possible to give an unproblematic treatment to time as a quantum object~\cite{AV08}; a rigorous definition for quantum reference frames~\cite{GS08}; whether it makes sense to define a quantum version of classical heat~\cite{GooldEtAl16}; the possibility of deriving quantum mechanics from informationally motivated laws~\cite{MM11}; quantum causal structures~\cite{ChavesElAl18} and the issue of quantum-to-classical transition~\cite{DCBM17} to mention but a few.  

Like all the foundational questions mentioned above, it is true that Wigner's original paradox (if any) has been turned inside out over and over again~\cite{BongEtAl20}. It is also true that a common sense is emerging in the field: only its many-party extensions deserve attention, as only they would add genuine quantum features to the analysis~\cite{LS14}. However, we strongly challenge this consensus. Before venturing into the complex contradictions of multipartite extensions, one must ensure that the foundational bipartite case is built on mathematically rigorous grounds. We argue that the original paradox remained conceptually murky precisely because a formal, operational definition of compatibility was missing. For this reason, we will use Wigner's original formulation to naturally reframe the incompatibility of agents' descriptions as an inference problem.

This `inferential turn' aligns with the research agenda advanced by refs.~\cite{LS14,Leifer14, Duarte20}. There, the authors see quantum theory as a noncommutative generalisation of Bayesian probability theory—one in which probability measures are replaced by density operators. In (standard) Bayesian probability theory, probabilities represent an agent's information, knowledge, or beliefs. This means that two different agents can naturally assign different probability distributions to the same random variable. Similarly, in approaches to quantum theory that view the quantum state as a representation of knowledge or belief, two agents can assign different states to the same quantum system---much like in Wigner's Thought Experiment. This raises the question: when are such different views compatible? In this sense, compatibility among distinct state assignments might help resolve the apparent paradox in Wigner's friend's thought experiment. A well-defined notion of compatibility within a Bayesian approach to probability theory, one that can help understand the apparent contradiction in Wigner's original formulation, is exactly what we will explore in this contribution.  

We have divided this contribution into two main parts. The first half addresses the original `paradox'. For completeness, we start by discussing Wigner's formulation in Sec.~\ref{Sec.WFP}, based on more modern references, namely~\cite{Cavalcanti21} and~\cite{LB21}. This Section aims to clarify that the `paradox' arises only from misconceptions about what (in)compatibility and agreement truly mean. By adopting a Bayesian perspective, we propose a sound definition of compatibility in Sec.~\ref{Sec.AgreementAndCompatibility}. In that Section, we also discuss the concept of agreement, an overlooked cornerstone of the thought experiment. Sec.~\ref{Sec.OurSolution} is where we elaborate on Wigner's original formulation and apply our concept of compatibility to conclude that there is nothing paradoxical in his \textit{Gedankenexperiment}. In the second half of the work, Sec. \ref{Sec.ouragreements} and Appendices~\ref{jointdistribution}, \ref{hybridstate} and \ref{ourpooling}, we turn to the means to achieve reconciliation between the agents' perspectives, that is, agreement. We present different reconciliation methods to accommodate subjective and objective interpretations of probability. We also draw attention to a derived case of interest by sprinkling a bit of benefit of the doubt into the studied paradox and by including the concept of quantum-state improvement \cite{LS14,Herbut2004}. In the aforementioned Appendices we briefly nod to two other solutions, which, in turn, are grounded in objectivist inference and the statistical pooling of quantum states \cite{LS14,Jacobs2002, Brun2015, Jacobs2005}. We conclude the paper by exploring further works, comparing our contribution with other approaches, and presenting a critical analysis of our own take on the subject.       

\section{Wigner's Friend Paradox}\label{Sec.WFP}

In this Section, we briefly review the original formulation of Wigner's Friend scenario and its pretence paradox.    

\subsection{Wigner's Original Formulation}\label{SubSec.WFPOriginal}

It revolves around two agents, namely Wigner (W) and his Friend (F)~\footnote{It may be the case that contemporary quantum information theorists may prefer the term `observers' to `agents'. We defend that the term `agent' is more appropriate because, whatever happens when one measures a quantum system, it cannot be regarded as a mere passive `observation'—in line with Eugene Wigner's original arguments~\cite{Wigner95}. The term `agent' also emphasises the agent-centric aspect deeply ingrained in the theoretical roots of subjectivist takes on probability theory~\cite{AA63,Savage1954,SEPDB22}.}. Wigner's Friend is locked in a laboratory and cannot communicate with Wigner. She [the Friend] will eventually interact with a qubit and will measure it afterwards. Wigner, on the other hand, is a superobserver and, in this hypothetical scenario, can use quantum theory to describe his Friend and the system she will interact with inside the hermetically closed laboratory. It is fundamental to note that both agents have agreed on every aspect of the experiment beforehand, including the fact that they know that each other knows how to use quantum theory appropriately. 

Wigner's Friend is locked in a fully working laboratory. The system (S) she interacts with is initially in a state represented by $\ket{0}_{S}$. It then undergoes a unitary Hadamard evolution, after which it is described by $\ket{+}_{S}=1/\sqrt{2}(\ket{0}_{S}+\ket{1}_{S})$. This is the point where the Friend measures the system $S$ with a projection-valued Measure (PVM) represented by $\Pi=\{\ketbra{0}{0}, \ketbra{1}{1}\}$. Given the Friend's knowledge of quantum theory, she knows that after her measurement the system ought to be described by $\ket{0}_{S}$ or by $\ket{1}_{S}$ depending on the measurement outcome and with equal probabilities obtained from Born's rule. The upshot of the Friend's perspective is that she collapses her description depending on the measurement outcome she experiences. 

Wigner sits outside the hypothetical laboratory. As a superobserver, he can describe both the system and the Friend with the formalism of quantum theory. In the thought experiment, he will describe his Friend (or his Friend's memory state) also as a two-level quantum system initially at $\ket{0}_{F}$—another fact agreed upon by both parties. Likewise, for Wigner, the system's initial state is represented by $\ket{0}_{S}$. System's and Friend's states jointly evolve, in Wigner's standpoint, as a composition of a Hadamard with a $CNOT$ controlled on $S$: 
\begin{align}
    \ket{00}_{SF} &\xmapsto{H \otimes \mathbb{1}} \frac{1}{\sqrt{2}}(\ket{0}_{S} + \ket{1}_{S})\otimes \ket{0}_{F} \nonumber, \\
 &\xmapsto{CNOT} \frac{1}{\sqrt{2}} \ket{00}_{SF} + \ket{11}_{SF} = \ket{\phi^{+}}_{SF} .
    \label{Eq.WignerDescriptionEvolution}
\end{align}
Wigner's ability to describe his Friend's and the system's states implies that he can measure either or both. Consequently, he can verify his predictions by performing, for example, the measurement described by $\Pi_{W}=\{\ketbra{\phi^{\pm}}{\phi^{\pm}}_{SF}, \ketbra{\psi^{\pm}}{\psi^{\pm}}_{SF} \}$. In this case, Wigner will obtain the outcome $\phi_{+}$ with probability 1, in an apparent contradiction with his Friend's observation. 

We can leverage the Friend's knowledge of quantum theory and turn the tables. The Friend may also want to describe Wigner's possible probabilities for the outcomes of the same measurement $\Pi_{W}=\{\ketbra{\phi^{\pm}}{\phi^{\pm}}_{SF}, \ketbra{\psi^{\pm}}{\psi^{\pm}}_{SF} \}$. According to the authors of ref.~\cite{LB21}, the Friend would, then, associate a probability 1/2 for the outcomes labelled by $\phi^{+}$ and $\phi^{-}$, in a yet another apparent contradiction, now with Wigner's predictions. We will return to this case in the next Sections.    

In sum, there appears to be an irreconcilable discrepancy. Wigner may swear that inside the laboratory, the joint state between his Friend and the system is given by $\ket{\phi^{+}}_{SF}$, whereas his Friend is sure that the joint state is either $\ket{00}_{SF}$ or $\ket{11}_{SF}$, each obtained with probability 1/2. Granted, in this version of the argument, we are allowing the Friend to measure herself, which is not strictly permitted by quantum theory due to the Heisenberg cut between what is measured and who is measuring. A rigorous resolution requires granting the Friend an external memory ($M$) to record her outcomes (e.g., recording $0$ as $\ket{0}_M$). While this introduces a bipartite target system $S \otimes M$, it does not smuggle in the multipartite contradictions typical of extended scenarios like the Frauchiger-Renner scenario~\cite{Frauchiger18}. Even if Wigner communicates beforehand his intention to treat the Friend ($F$) as a quantum system (e.g., a two-level system in $\mathbb{C}^2$) subject to unitary evolution, the Friend does not view herself as such. From her perspective, the most she could accept is that what Wigner labels as the Friend ($F$) actually refers to the physical record made by the memory apparatus ($M$) entangled with the quantum system ($S$). The lab, to Wigner, remains structurally bipartite ($S \otimes M$), not tripartite ($S \otimes M \otimes F$). The agent $F$ remains strictly an epistemic entity outside the Heisenberg cut, preventing self-reference issues without generating contradictions. We deliberately restrict our focus to the original two-epistemic-agent scenario to establish our Bayesian formalism. Applying this framework to evaluate many-party extensions is a natural next step, but it is left for future work. Wigner and the Friend are, now, interested in describing the Friend's external memory and the quantum system. To simplify, we will continue referring to M by the index F. Regardless of whose standpoint we assume, the naive upshot is that Wigner's and his Friend's perspectives `paradoxically' clash with one another. We may let Eugene Wigner speak for himself~\footnote{See~\cite{Wigner95} p. 180.}:
\begin{center}
\emph{``This is a contradiction, because the state described by the wave function [$\ket{\phi^{+}}_{SF}$] describes a state that has properties that neither $[\ket{00}_{SF}]$ nor $[\ket{11}_{SF}]$ has''}. 
\end{center}
The fact of the matter that puzzled Eugene Wigner in~\cite{Wigner95}, and many others after him, is that if quantum theory is valid at all scales, then Wigner's description of the thought experiment would have to be compatible with Wigner's Friend's experience of obtaining a definite outcome inside her laboratory. Presumably, for Eugene Wigner, it is precisely this lack of compatibility that should be taken into account---which, then, sets the concept as a cornerstone of the `paradox'~\footnote{Compatibility is just an ingredient. Agreement, knowledge, belief, knowledge of knowledge, knowledge of knowledge of knowledge, group knowledge, common knowledge, and the like are also fundamental to this scenario. What we advance in this contribution is that we are flooded with paradoxical interpretations of Eugene Wigner's thought experiment precisely because these crucial concepts have been consistently overlooked. Granted, we only work out the role of a sound definition of 'compatibility' within a Bayesian inference context, but this seems more than enough for a first step toward getting rid of superficial takes on the paradox. Other authors are also walking on the same path—see refs.~\cite{HC23} and~\cite{VN19} }. 

In the next Sections, we will make the role played by compatibility in Eugene Wigner's thought experiment more transparent. Before that, we take a brief detour to review other sophisticated resolutions and possible extensions of the pretence paradox. 

\subsection{Extended Wigner's Friend Scenarios and Possible Resolutions}

Even though the Wigner's Friend Experiment---as originally proposed in Wigner's historical contribution to the mind and body debate~\cite{Wigner95}---can certainly be regarded as a tension between unitary evolution and the measurement 'collapse' in quantum theory, there have been a proliferation of arguments about the implications of the thought experiment to quantum theory~\cite{Brukner20}. We have seen many resolutions of its allegedly paradoxical conclusions: from alternative interpretations to quantum mechanics~\cite{SYL23} to toy models that ultimately show that what we would definitely see as classical physics may also exhibit the same contradictions at the heart of Wigner's imagined scenario~\cite{LB21, JM24}.

Aligned with the perspective we want to advance in the present work, Brukner~\cite{brukner2015} addresses the ambiguity of the measurement problem by arguing against the existence of \emph{facts of the world per se}. To avoid the theoretical pitfalls of hidden-variable programs when dealing with Wigner's and his friend's coexisting records, he proposes that measurement records and facts can exist only relative to the observer. Taking a further step, Baumann and Brukner~\cite{baumann2019} reframed the paradox as an inferential approach,  by treating the friend strictly as a rational agent. They demonstrate that the apparent contradiction is bypassed when the friend realises that conditioning her predictions solely on events inside her isolated laboratory is irrational--- maybe due to what we call \emph{Wigner stubbornness} as will be mentioned in \ref{cromwell}. Reconciliation is achieved by allowing the friend to extend her conditioning to information available outside the lab, effectively enabling her to adopt Wigner's perspective alongside her own through a modified Born rule. More recently, Del Santo, Manzano, and Brukner~\cite{delsanto2025} formalised this informational divide through the concept of \emph{information bubbles} to claim that there is no paradox without necessarily asserting that one of the agents should supposedly be taken as the correct observer to the detriment of an error in the other's worldview. Rather than granting Wigner the \emph{absolutely correct} state description due to his superobserver status, they advocate for a \emph{relative objectivity}, where state assignments are objective only relative to a specific bubble.

Against the background of such a thorough study of Wigner's original \textit{Gedankenexperiment}, it is now believed that only extended versions of his original formulation present a challenge to classically ingrained misconceptions of quantum theory~\cite{LB21}. A recent paper by K. W. Bong et al. advances such an extended framework and puts forward a local friendliness hypothesis, which is supposed to be violated in certain quantum setups~\cite{BongEtAl20}. Bong's paper is the most prominent example of those extended scenarios~\cite{SYL23}. 

Bringing our own perspective to the discussion, we argue that the problem may lie in the lack of a sound definition of what compatibility and agreement should truly mean in a Bayesian context, and in the pursuit of agreement where mere compatibility is sufficient. In this sense, although Wigner's historical contribution has been turned inside out repeatedly, we believe there is more to it and that the thought experiment can shed light on how we see quantum theory---and, in particular, its presumed parallel with classical probability theory. 
Furthermore, we are interested in exploring how our resolution of the paradox behaves when a twist of the benefit of the doubt is added into the scenario, as discussed in Sec.~\ref {SubSec.OurSolution} and Sec.~\ref{benefitofdoubt}.

So next, we will explore how the notions of compatibility and agreement are fundamental to resolving the paradox once and for all.  

%
\section{COMPATIBILITY AND AGREEMENT}\label{Sec.AgreementAndCompatibility}
Every analysis of Eugene Wigner's thought experiment has a central aspect in common. Almost all the takes on Wigner's hypothetical scenario elaborates on the possibility of conflicting descriptions between two or more agents about what is regarded as the same reality. In other words,  down to the core of those analyses is the fact that they depend, directly or indirectly, on agent-centred notions that ought to be made more explicit: agreement and compatibility~\footnote{And rational behaviour, for that matter.}.

For the 'paradox' to surface, Wigner and his Friend must agree on the use of quantum mechanics, on how to estimate probabilities via the Born rule, on the correct time in which the Friend will measure the system, on each other's knowledge of each other's knowledge of quantum theory~\footnote{The repetition here is intentional. References~\cite{LD22} and~\cite{ContrerasEtAl21} elaborate on this topic.}, and so on. Likewise, the `paradoxical' conclusion relies on the apparent incompatibility between Wigner's and his Friend's descriptions. It is tacitly assumed that once their descriptions do not exactly match one another, there must be something deeply wrong with either party involved in the thought experiment.    

We challenge this view. We advocate that Wigner's Friend Paradox is a problem that should be looked at through the lenses of (Bayesian) inference—in the end, it is nothing but the problem of two agents trying to infer what is happening in a particular situation by putting themselves in each other's shoes. By doing so, we are led directly to well-motivated concepts of compatibility, improvement, and pooling~\cite{LS14}. We will show that the usual paradox withers away under these inferential notions.  

\subsection{Compatibility in Classical Probability Theory}\label{SubSec.ClassicalCompatibility}

We first consider compatibility for classical random variables. From a Bayesian standpoint, there are two ways in which probabilistic descriptions of random variables may differ across multiple agents: one may be an objectivist or a subjectivist. Although we address each case separately, we will see that their mathematical characterisation is, nonetheless, the same. 

\subsubsection{Objective Bayesian Compatibility}\label{SubSubSec.ClassicalObjectivist}

For the objective Bayesian, there is only one way in which the probability assignment of two or more agents may differ. They must have been exposed to different datasets. Adhering to ref.~\cite{LS14}, we name the agents Wanda (W) and Freddy (F), and the motivation for (objective) compatibility is as follows.

Consider the situation in which the agents want to decide on the potential differences in their descriptions of a random variable $Y$. From an objectivist standpoint, there is only one explanation: at some point in the past, Wanda learned the value of a random variable $X_1$, and Freddy learned the value of another random variable $X_2$. According to this view, there must be a unique probability distribution $\mathbb{P}(Y,X_1,X_2)$ that Wanda and Freddy ought to agree upon and assign to those three random variables before they have observed the values of $X_1$ and $X_2$~\footnote{Agreement will be discussed in the next Section, but one should notice how it cropped out and is central already at this stage of the discussion.}. Wanda's and Freddy's prior distributions for $Y$ coincide and are simply the marginal $\mathbb{P}(Y)=\sum_{X_1,X_2}\mathbb{P}(Y,X_1,X_2)$. Upon experiencing a definite value $x_i$ of $X_i$, Wanda and Freddy update their assignments to their posteriors
\begin{align}
    \mathbb{P}_{W}(Y) = \mathbb{P}_{W}(Y | X_1 = x_1) \mbox{ for Wanda} ,
\end{align}
\begin{align}
    \mathbb{P}_{F}(Y) = \mathbb{P}_{F}(Y | X_2 = x_2) \mbox{ for Freddy}.
\end{align}
This is how an objective Bayesian would describe the reason why Wanda and Freddy arrived at the two possibly different, but not so discrepant, descriptions $\mathbb{P}_{W}(Y)$ and $\mathbb{P}_{F}(Y)$ for the same random variable $Y$. 

In this sense, the objective Bayesian would adopt the following definition for compatibility.
\begin{definition}[Classical Objective Bayesian Compatibility]
Two probability distributions $\mathbb{P}_{W}(Y)$ and $\mathbb{P}_{F}(Y)$ are \emph{compatible} whenever it is possible to find a pair of random variables $X_1$ and $X_2$, a joint probability distribution $\mathbb{P}(Y,X_1,X_2)$, and a pair of outcomes $(x_1,x_2) \in \mbox{Out}(X_1) \times \mbox{Out}(X_2)$ such that:
\begin{enumerate}
    \item [(a)] $\mathbb{P}(X_1=x_1,X_2=x_2) \neq 0$.
    \item [(b)] $\mathbb{P}_{W}(Y) = \mathbb{P}(Y | X_1 = x_1)$.
    \item [(c)] $\mathbb{P}_{F}(Y) = \mathbb{P}(Y | X_2 = x_2)$.
\end{enumerate}
\label{Def.ClassicalCompatibilityObjective}
\end{definition}
What is remarkable is that the definition above can be promptly checked by comparing the supports of $\mathbb{P}_{W}(Y)$ and $\mathbb{P}_{F}(Y)$.

\begin{theorem}
Two distributions $\mathbb{P}_{W}(Y)$ and $\mathbb{P}_{F}(Y)$ are compatible in the objective Bayesian sense if, and only if, the intersection of their support is non-trivial:
\begin{align}
    \mbox{supp}(\mathbb{P}_{W}(Y)) \cap \mbox{supp}(\mathbb{P}_{F}(Y)) \neq \emptyset.
    \label{Eq.ThmClassicalCompatibilityObjective}
\end{align}
\label{Thm.ClassicalCompatibilityObjective}
\end{theorem}
We emphasise that the Def.~\ref{Def.ClassicalCompatibilityObjective} comes with two critical assumptions. First, it demands to be possible to find a joint probability distribution for all the involved random variables, such that the correct marginals are secured. A property we know is not always true for every collection of random variables~\cite{AA63}. Secondly, it is also assumed that both outcomes $x_1$ and $x_2$ can be simultaneously observed with non-zero probability, and that both agents agree on that in advance—even if they are kept far away from each other; an issue that plagues other situations involving agreement between two or more agents~\cite{Demey14}.

All in all, the objective Bayesian approach requires the existence of a joint probability distribution $\mathbb{P}(Y,X_1,X_2)$. This distribution represents the collective description of the random variable $Y$ the agents want to describe,  together with the random variables $X_1$ and $X_2$ representing their access to different pieces of information. From a subjective Bayesian standpoint, this is not a necessity. 

\subsubsection{Subjective Bayesian Compatibility}\label{SubSubSec.ClassicalSubjectivist}

Subjective Bayesians do not rule out the case in which agents start out from different readings of the world~\cite{SEPBayesianEpistemology24}. In this sense, there is no reason why the difference between the agents's assignments, $\mathbb{P}_{W}(Y)$ and $\mathbb{P}_{F}(Y)$, has to come from the agents's access to different data. Consequently, definition~\ref{Def.ClassicalCompatibilityObjective} is not appropriate to this case. What can be done in this situation?

Because subjective Bayesians do not rationalise how agents end up with their discrepant descriptions, what is at stake in this standpoint is whether these descriptions can be reconciled in the future or if their discrepancy is so substantial that the agents ought to make a wholesome revision of their views to agree with each other~\footnote{Using the qualifier 'substantial' emphasises cases where there is no compatibility between the agents. The definition we work with in this contribution does not allow for quantifying the lack of compatibility between the agents. Smoother measures of incompatibility will be investigated in future works.}. A rule for deciding whether such reconciliation is possible is the following. 
\begin{definition}[Classical Subjective Bayesian Compatibility]
    Two probability distributions $\mathbb{P}_{W}(Y)$ and $\mathbb{P}_{F}(Y)$ are \emph{compatible} whenever it is possible to find a random variable $X$ and a conditional probability $\mathbb{P}(X|Y)$ such that there exists a particular $x' \in \mbox{Out}(X)$ such that
    \begin{enumerate}
        \item [(a)] $\sum_{y \in \mbox{Out}(Y)}\mathbb{P}(X=x' | Y=y) \mathbb{P}_{J}(Y=y) \neq 0$.
        \item[(b)] $\mathbb{P}_{W}(Y =y | X=x') = \mathbb{P}_{F}(Y =y | X=x'), \,\, \forall \,\, y$.
    \end{enumerate}
    Where the conditional assignment in item (b) is defined, for $J \in \{W,F\}$, as: 
    \begin{align}
        \mathbb{P}_{J}(Y=y | X=x) := \frac{\mathbb{P}(X=x | Y=y) \mathbb{P}_{J}(Y=y)}{\sum_{y'} \mathbb{P}(X=x | Y' = y') \mathbb{P}_{J}(Y' = y') }.
    \end{align}
    \label{Def.ClassicalSubjectiveCompatibility}
\end{definition}
This definition says that Wanda's and Freddy's probabilistic assignments are compatible (from a subjective Bayesian standpoint) when both agents agree on an experiment represented by a random variable $X$ as well as on a likelihood function $\mathbb{P}(X | Y)$ for that experiment such that experiencing a particular outcome, $x' \in \mbox{Out}(X)$, makes Wanda and Freddy update and assign identical conditional probabilities $\mathbb{P}_{W}(Y =y | X=x') = \mathbb{P}_{F}(Y =y | X=x')$. In other words, compatibility is translated into agreement in another experiment that will decide whether or not the agents' assignments are substantially discrepant. 

Whereas compatibility in the objective sense looks to the past to explain and reconcile the differences among the agents' probabilistic assignments, the notion of compatibility in the subjective sense is more concerned with whether the agents' assignments can be reconciled in the future. Their content is, consequently, starkly and naturally different. Nonetheless, the next theorem shows that both the objective and subjective compatibility notions are equivalent to exactly the same mathematical condition:  we only have to look at the support of $\mathbb{P}_{W}(Y)$ and $\mathbb{P}_{F}(Y)$ to determine their compatibility.
\begin{theorem}
Two distributions $\mathbb{P}_{W}(Y)$ and $\mathbb{P}_{F}(Y)$ are compatible in the subjective Bayesian sense if, and only if, the intersection of their support is non-trivial:
\begin{align}
    \mbox{supp}(\mathbb{P}_{W}(Y)) \cap \mbox{supp}(\mathbb{P}_{F}(Y)) \neq \emptyset.
\label{Eq.ThmClassicalCompatibilitySubjective}
\end{align}
\label{Thm.ClassicalCompatibilitySubjective}
\end{theorem}
It is out of the scope of this contribution to demonstrate ~\cref{Thm.ClassicalCompatibilityObjective} and~\cref{Thm.ClassicalCompatibilitySubjective}. Here, we will use the characterisation and interpretation they provide to shed light on Eugene Wigner's alleged paradox. A detailed proof for those theorems can be found in ref.~\cite{LS14}. In that same work, the authors investigate a generalisation of both compatibility notions, but in quantum theory. We will briefly review their main results in the next Section, as they will also be used in our argument against paradoxical takes on Eugene Wigner's Thought Experiment.  

\subsection{Compatibility in Quantum Theory}\label{SubSec.QuantumCompatibility}

Strengthening the parallel between classical probability theory and quantum theory, in ref.~\cite{LS14}, the authors extended the notion of compatibility for classical agents to compatibility for agents whose probabilistic assignment is quantum. Their definition of (quantum) compatibility is based on the formalism of conditional quantum states developed in~\cite{LS13} and mimics its classical counterpart, including the characterisation theorems~\ref{Thm.ClassicalCompatibilityObjective} and~\ref{Thm.ClassicalCompatibilitySubjective}---yet another natural consequence of adopting the quantum conditional states approach to reasoning about the quantum world  

We defer to Appendix~\ref{SubApp.DefQuantumCompatibility} the definitions of (quantum) compatibility in the subjective and objective senses as borrowed from ref~\cite{LS14}. As we alluded to before, they are heavily based on the quantum conditional states formalism and will not review it here — it is completely out of scope for this contribution. To cut straight to the chase, we will assume the definitions' associated characterisation theorems as the correct definitions for (quantum) compatibility and use them to uncloud Wigner's 'paradox' in the next Section. The quantum compatibility definitions (Defs.~\ref{Def.QuantumObjectiveCompatibility} and~\ref{Def.QuantumSubjectiveCompatibility}) holds whenever the BFM compatibility condition~\cite{Brun2002}
is satisfied, that is:
\begin{theorem}[Characterisation Theorem for Quantum Compatibility]
Two states $\sigma_{S}^{W}$ and $\sigma_{S}^{F}$ are compatible if and only if the intersection of their supports is not trivial. In other words, they are compatible iff:
\begin{align}
    \mbox{supp}(\sigma_{S}^{W}) \cap \mbox{supp}(\sigma_{S}^{F}) \neq \{\vec{0}\}.
    \label{Eq.ThmQuantumCompatibility}
\end{align}
    \label{Thm.QuantumCompatibility}
\end{theorem}
\textbf{Remark:} Recall that the support of a state $\rho$ is the subspace spanned by its eigenvectors associated with non-null eigenvalues. Consequently, the null vector always belongs to $\mbox{supp}(\rho)$. If Eq.~\eqref{Eq.ThmQuantumCompatibility} above was more similar to its classical counterparts, Eqs.~\eqref{Eq.ThmClassicalCompatibilityObjective} and~\eqref{Eq.ThmClassicalCompatibilitySubjective}, we would never find two incompatible quantum assignments. It is remarkable how the quantum compatibility definitions (Defs.~\ref{Def.QuantumObjectiveCompatibility} and~\ref{Def.QuantumSubjectiveCompatibility} ) mimic their classical counterparts, generating essentially the same characterisation theorem but not exactly the same to the point of trivialising the entire concept.   

It is also crucial to dispel a common misconception: quantum compatibility is \emph{not} a mere restatement of non-orthogonality. While it is true that orthogonal states are strictly incompatible (their supports are orthogonal subspaces, intersecting only at the origin), the converse does not hold. Not all non-orthogonal states are compatible. Consider, for instance, two agents assigning distinct pure states, such as $\rho = \vert{}0\rangle\langle0\vert{}$ and $\sigma = \vert{}+\rangle\langle+\vert{}$. These states are clearly non-orthogonal ($\mbox{Tr}(\rho\sigma) = 1/2 > 0$). However, their supports are distinct one-dimensional rays ($\operatorname{span}\{\vert{}0\rangle\}$ and $\operatorname{span}\{\vert{}+\rangle\}$), whose intersection is strictly trivial ($\{\vec{0}\}$). Hence, they are fundamentally incompatible. Compatibility is a significantly more restrictive requirement than non-orthogonality. The subsequent sections will show that the reconciliation of Wigner's and his Friend's descriptions—whether by constructing an objectivist joint state or a subjectivist likelihood experiment—is mathematically grounded entirely on this non-trivial intersection. If their states were merely non-orthogonal but incompatible, the existence of such reconciliation structures would be mathematically impossible.

Let us sum up what we have so far. Two classical probability assignments are compatible whenever their supports intersect nontrivially. Similarly, two quantum probability assignments are compatible whenever their supports intersect nontrivially. As a result, on the one hand, we can consider Wigner (Wanda) and his Friend (Freddy) as classical agents (rational beings assigning classical probability distributions to events) and call upon the classical definition to decide whether their assigned probabilities are incompatible. On the other hand, viewing Wigner and the Friend as quantum agents (as the paradox intends), we can compare their quantum assignments and decide, with the aid of Thm.~\ref{Thm.QuantumCompatibility}, whether they are incompatible. We will see in a minute why this resolves any remaining questions about the discrepancy in Wigner's and his Friend's assignments. There is, though, a connected notion that ought to be explored if we want to light up the remaining obscure corners in Eugene Wigner's original thought experiment. Wigner and his Friend must agree on certain aspects of the experiment beforehand.

\subsection{Agreement}\label{SubSec.Agreement}
Although compatibility may be readily turned into a rigorous notion—readily usable in Wigner's thought experiment—agreement still remains a subtler notion. There is a substantial body of literature about reaching agreement; either in classical~\cite{Degroot74, Demey14, Samet22, RW90}, in quantum~\cite{ContrerasEtAl21} or in more general probability theories~\cite{LD22}. In this contribution, we only scratch the surface of the concept. We do not want to develop a protocol where Wigner cannot agree to disagree with his Friend~\cite{Aumann76}. This is not the intention here. What we want is way more modest. 

Recall that, in this contribution, we want to explore contentious parts of Eugene Wigner's original formulation.  In particular, it should be stressed that, in the thought experiment, there are several things both parties must agree on; otherwise, Wigner's and the Friend's descriptions would naturally be incompatible (more on that in a minute).

Contrary to what it may sound like, this is not dull work. Not least, as we just saw, agreement is at the heart of the subjective notion of compatibility. Suppose agents could not agree on the existence of a second random variable's probability distribution, one that would resolve the differences in the agents' descriptions—see Def.~\ref{Def.ClassicalSubjectiveCompatibility}. In that case, there is no hope of deciding whether the two worldviews are compatible, at least not under a subjective Bayesian standpoint. As we will see in the next Sections, our analysis goes beyond that point. It will become more evident that agreement between the agents lurks behind every corner of Eugene Wigner's original formulation, and the lack of it can be explored to ruin the entire edifice on which the `paradox` is built.~\cite{Aaronson04}. Nonetheless, for concreteness, we should at least sketch a working definition of what we mean by agreement in this work.

Our notion of agreement is inspired by Williams's idea of a thick notion~\cite{SepThickNotion}. We say that two or more agents \emph{agree} on something when their descriptions of that thing are exactly the same at some point in time; and we will stick to this notion in both its classical and quantum forms. More precisely:
\begin{definition}[Classical Agreement]\label{Def.ThickNotionAgreementClassical}
   We say that the agents $A_1,A_2, \cdots ,A_N$ agree on a random variable $X$ whenever 
    \begin{align}
        \mathbb{P}_{A_1}(X) = \mathbb{P}_{A_2}(X) = \cdots = \mathbb{P}_{A_N}(X),   
        \label{Eq.DefThickNotionAgreementClassical}
    \end{align}
that is, when their probability distributions for that particular random variable coincide with each other.  
\end{definition}
\begin{definition}[Quantum Agreement]\label{Def.ThickNotionAgreementQuantum}
   We say that the agents $A_1,A_2, \cdots ,A_N$ agree on a random variable $X$ whenever 
    \begin{align}
        \rho^{A_1}_{X} = \rho^{A_2}_{X} = \cdots = \rho^{A_N}_{X},   
        \label{Eq.DefThickNotionAgreementQuantum}
    \end{align}
that is, when the (conditional) states that each agent assigns to that particular random variable coincide with each other.  
\end{definition}
Consequently, for example, we say that Wigner and his Friend agree on the initial state of the quantum system inside the laboratory when either agent assigns the same $\ket{0}_{S}$ to it. We also say that there is an agreement on the instant of time in which the Friend will run her measurement; and this amounts to $\mathbb{P}_{\mbox{Friend}}(T_{M}=t) = \mathbb{P}_{\mbox{Wigner}}(T_{M}=t)$, where $T_{M}$ is the random variable describing the possible instants of time in which Wigner's Friend's measurement could have been performed. 

We emphasise that far subtler types of agreements are also possible within this definition, and that we are not working with the concept of certainty, but only (probabilistic) agreement between the agents—for a more elaborate discussion, see~\cite{ContrerasEtAl21} and~\cite{LD22}. In this sense, even when agents are uncertain about the outcome of an experiment, they may still agree with each other. We also assume that Wigner and Wigner's Friend agree on each other's knowledge of how to use quantum theory. 

As said before, in the next Section we explore how the notion of agreement seems to be latent in every corner of Eugene Wigner's thought experiment and how an absence of agreement can lead to clashing descriptions between Wigner and his friend. Presumably, these clashing descriptions would not be viewed as a `paradoxical' situation, as they originate from totally incongruent descriptions. But these examples are motivating. Recall that incompatibility precludes agreement, and, in this light, we advocate that there is no paradox whatsoever in the original thought experiment. The pretence paradox only arises when we seek agreement where demanding compatibility is more than enough for the agents' descriptions—at least when we reframe the question within a (Bayesian) inference problem. 

\section{What is paradoxical in the paradox?}\label{Sec.OurSolution}

In its original formulation, Wigner's thought experiment was intended to show an irreconcilable and `paradoxical' discrepancy between Wigner's and his Friend's descriptions~\cite{Wigner95}. The alleged incompatibility between the two agents' descriptions would be crucial to prove the pivot role played by agents' conscious decisions in quantum mechanics—and, more generally, in physics. What is missed from Eugene Wigner's original analysis (and several subsequent ones) is a sound definition of what incompatibility truly is. We believe that this notion is fundamental not because of a taste for abstraction, but because it is precisely this concept that gives rise to the `paradox' of Eugene Wigner's thought experiment. We have made this concept precise in the previous Sections, and now we argue that we can dispense with the `paradox'.

\subsection{Classical Assignment}\label{SubSec.OurSolutionClassical}

On the one hand, consider the statistics arising from Wigner's hypothetical measurements. Recall, Wigner's measurement is described by $\Pi_{W}=\{\ketbra{\phi^{\pm}}{\phi^{\pm}}_{SF}, \ketbra{\psi^{\pm}}{\psi^{\pm}}_{SF} \}$ and the state he assigns to the compound system $SF$ inside the laboratory is $\ketbra{\phi^{+}}{\phi^{+}}_{SF}$. In this case, if Wigner knows how to use quantum theory, Born's rule guarantees that his probability assignment is given by: 
\begin{align}
    \mathbb{P}_{W}(\mathcal{M}) =(1,0,0,0),
    \label{Eq.ClassicalSolutionWignersAssignment}
\end{align}
where, 
\begin{align}
    \mbox{Prob}( \phi_{+} \,\, | \,\, \ketbra{\phi^{+}}{\phi^{+}}_{SF})&=1 \nonumber  ,\\
    \mbox{Prob}( \phi_{-} \,\, | \,\, \ketbra{\phi^{+}}{\phi^{+}}_{SF})&=0 , \\
    \mbox{Prob}( \psi_{+} \,\, | \,\, \ketbra{\phi^{+}}{\phi^{+}}_{SF})&=0 \nonumber , \\
    \mbox{Prob}( \psi_{-} \,\, | \,\, \ketbra{\phi^{+}}{\phi^{+}}_{SF})&=0, \nonumber
\end{align}
and $\mathcal{M}$ represents the measurement determined by the PVM $\Pi_{W}$.

Now, on the other hand, assuming that Wigner's Friend agrees with Wigner on what is being measured by her and that she also knows how to use quantum theory, her probability assignment is:
\begin{align}
    \mathbb{P}_{F}(\mathcal{M}) =\left(\frac{1}{2},\frac{1}{2},0,0\right),
    \label{Eq.ClassicalSolutionFriendsAssignment}
\end{align}
where, 
\begin{align}
    \mbox{Prob}\left( \phi_{+} \,\, | \,\, \frac{1}{2}\ketbra{00}{00}+\frac{1}{2}\ketbra{11}{11} \right)&=1/2 \nonumber , \\
    \mbox{Prob}\left( \phi_{-} \,\, | \,\, \frac{1}{2}\ketbra{00}{00}+\frac{1}{2}\ketbra{11}{11} \right)&=1/2  ,\\
    \mbox{Prob}\left( \psi_{+} \,\, | \,\, \frac{1}{2}\ketbra{00}{00}+\frac{1}{2}\ketbra{11}{11} \right)&=0 \nonumber , \\
    \mbox{Prob}\left( \psi_{-} \,\, | \,\, \frac{1}{2}\ketbra{00}{00}+\frac{1}{2}\ketbra{11}{11} \right)&=0. \nonumber
\end{align}
Assuming that Wigner and his Friend reason like rational Bayesian agents, they will settle the possible incompatibility of their probability assignments according to the processes we described before in Sec.~\ref{SubSec.ClassicalCompatibility}. They may want to use Def.~\ref{Def.ClassicalCompatibilityObjective} if they are objective Bayesians or Def.~\ref{Def.ClassicalSubjectiveCompatibility} if they are subjective Bayesians. Regardless of their standpoint, they will settle their possible dispute by comparing the support of their distributions—they only need to check Eqs.~\eqref{Eq.ThmClassicalCompatibilityObjective} and~\eqref{Eq.ThmClassicalCompatibilitySubjective}. 

Because $\mbox{supp}(\mathbb{P}_{W}(\mathcal{M})) \cap \mbox{supp}(\mathbb{P}_{F}(\mathcal{M})) \neq \emptyset$ their assignments \emph{are compatible}. There is no incompatibility and, therefore, no paradoxical scenario. The agents simply had access to different experimental data (objective) or had been through different experiences (subjective), which corresponds precisely to the situation imagined in Eugene Wigner's Original Thought Experiment. 

In sum, if we reframe Eugene Wigner's Original thought experiment as an inference problem where the involved agents' reasoning is modelled within a (classical) Bayesian perspective, the apparent incompatibility between Wigner's and his Friend's probabilistic descriptions is naturally explained, and the paradox withers away. What if we considered quantum theory as a theory of quantum probabilistic assignments?  

\subsection{Quantum Assignment}\label{SubSec.OurSolutionQuantum}

Let us consider quantum theory as a probabilistic theory for a moment. This idea was previously alluded to in Sec.~\ref{Sec.AgreementAndCompatibility}, following the reasoning of refs.~\cite{Leifer06,LS13,LS14, Duarte20}. Within such a framework, (quantum) agents assign quantum conditional states to regions of interest, in analogy to what classical agents do for random variables. Agents' probabilistic assignments are carried over from one region of interest to another via belief propagation, also resembling its classical counterpart. 
We can think of concepts such as the Born rule, ensemble averaging, quantum map actions, channel composition, and the non-selective update rule as instances of belief propagation from one region to another. Table 2 in reference~\cite{LS13} provides a more visual representation of this connection. It is this aspect of the formalism that we use in this Section.

From inside the laboratory, the Friend's description of what has been stored in her memory, given what she has experienced, is given by the following assigned state~\footnote{Recall the above discussion about the Friend being able to describe herself via quantum theory.}:
\begin{align}
    \rho_{X}^{F} = \mbox{Tr}_{SF} \left[ \varrho_{X | SF}^{F} \rho_{SF} \right] = \begin{pmatrix}
\frac{1}2 & 0 & 0 & 0 \\
0         & \frac{1}{2} & 0 & 0 \\
0         & 0 & 0 & 0 \\
0         & 0 & 0 & 0 
\end{pmatrix}.
\end{align}
$X$ represents the classical region associated with the Friend's measurement. Similarly, $SF$ is the compound region consisting of the Friend's recording memory and the evolving system. 

It is important to emphasize that $\rho_{X}^{F}$ represents the classical probability distribution over the measurement outcomes and it is expressed in the Bell basis. This clarifies why the matrix is $\operatorname{diag}(1/2, 1/2, 0, 0)$ rather than $\operatorname{diag}(1/2, 0, 0, 1/2)$, which would be the representation of the Friend's quantum state $\sigma_F = \frac{1}{2}\vert{}00\rangle\langle00\vert{} + \frac{1}{2}\vert{}11\rangle\langle11\vert{}$ in the computational ($Z$) basis. Similarly, $SF$ is the compound region consisting of the Friend's recording memory and the evolving system.

For Wigner, the situation is slightly more elaborate. Because there is no need to refer to the measurement statistics to compare Wigner's and his Friend's assignments, the formalism of conditional states allows for at least two related situations. We may want to $(i)$ consider the case in which Wigner intends to perform the same PVM as before $\Pi_{W}=\{\ketbra{\phi^{\pm}}{\phi^{\pm}}_{SF}, \ketbra{\psi^{\pm}}{\psi^{\pm}}_{SF} \}$, with the four classical outcomes $\{\phi^{+}, \phi^{-}, \psi^{+}, \psi^{-}\}$---and in this case we are bound to look to the measurement statistics---or we may say $(ii)$ that his assignment simply consists of the state arising from joint evolution between the Friend and the quantum system inside the laboratory. 

When there is a measurement, belief propagation~\cite{LS13,LS14} can be used to derive Wigner's assignment:
\begin{align}
    \rho_{X}^{W} = \mbox{Tr}_{SF} \left[ \varrho_{X | S'F'}^{W} \varrho_{S'F' | SF}^{W} \rho_{SF} \right] = \begin{pmatrix}
1 & 0 & 0 & 0 \\
0         & 0 & 0 & 0 \\
0         & 0 & 0 & 0 \\
0         & 0 & 0 & 0 
\end{pmatrix}.
\label{Eq.WignerQuantumAssignmentMeasurement}
\end{align}
Similarly, we can use belief propagation~\cite{LS13, LS14} to obtain Wigner's assignment in the case he stops short of the measurement:
\begin{align}
    \rho_{S'F'}^{W} = \mbox{Tr}_{SF} \left[ \varrho_{S'F' | SF}^{W} \rho_{SF} \right] = \ketbra{\phi^{+}}{\phi^{+}}.
\end{align}
In either case, $SF$ and its primed version $S'F'$ are the compound regions representing the actual sites where the joint evolution occurs. In equation~\eqref{Eq.WignerQuantumAssignmentMeasurement}, the sub-index $X$ represents the classical region associated with Wigner's measurement. 

Now we can compare Wigner's and his Friend's (quantum) assignments and decide whether they are incompatible---the cornerstone of Eugene Wigner's `paradox'. According to thm.~\ref{Thm.QuantumCompatibility}, to determine the compatibility between the (quantum) agents' probabilistic assignments, we only need to examine the supports of $\rho_{S'F'}^{W}$, $\rho_{X}^{W}$, and $\rho_{X}^{F}$. Not surprisingly, it turns out that:
\begin{align}
    \mbox{supp}(\rho_{X}^{F}) \cap \mbox{supp}(\rho_{X}^{W}) &\neq \{\vec{0} \}
    \label{Eq.QuantumAssignmentsAreCompatible1}
\end{align}
and
\begin{align}
    \mbox{supp}(\rho_{X}^{F}) \cap \mbox{supp}(\rho_{S'F'}^{W}) &\neq \{ \vec{0}\}.
    \label{Eq.QuantumAssignmentsAreCompatible2}
\end{align}
Therefore, by reinterpreting Eugene Wigner's Thought Experiment as a problem of Bayesian inference, we can conclude that there is no conflict between Wigner's and his Friend's assessments. Interestingly, this conclusion holds true whether they are viewed as classical or quantum agents.

This argument shows that by removing incompatibility from the equation, we are left with no paradoxical interpretations of the historical formulation of Eugene Wigner's thought experiment. The difference between Wigner's and his Friend's descriptions is solely due to the Friend having access to data or experiences that Wigner did not have access to. If both agents are subjective Bayesians after the laboratory is open, they can agree on another experiment to resolve their dispute (thms.~\ref{Thm.ClassicalCompatibilitySubjective} and~\ref{Thm.QuantumCompatibility}). If both agents are objective Bayesians, after the laboratory is open, they can share the data they had access to (mainly the Friend in this case) and then update their probabilistic assignments accordingly (thms.~\ref{Thm.ClassicalCompatibilityObjective} and~\ref{Thm.QuantumCompatibility}). In either case, there is no incompatibility, and there is no paradox. The agents cannot agree to disagree. 

\subsection{The role played by agreement and compatibility in Eugene Wigner's Thought Experiment}\label{SubSec.OurSolution}
The fact that Wigner and his friend cannot agree to disagree should be clearer now. 
However, this might not be the case if they disagreed on several hidden aspects of the thought experiment's setup.

\subsubsection{Wigner assigns the wrong initial state}

To begin with, consider the case where Wigner assigns $\ket{11}_{SF}$ for the initial state shared by the Friend and the quantum system inside the laboratory. System's and Friend's states still jointly evolve, from Wigner's standpoint, as a composition of a Hadamard with a $CNOT$ controlled on $S$. 
\begin{align}
    \ket{11}_{SF} &\xmapsto{H \otimes \mathbb{1}} \frac{1}{\sqrt{2}}(\ket{0}_{S} - \ket{1}_{S})\otimes \ket{1}_{F} \nonumber, \\
 &\xmapsto{CNOT} \frac{1}{\sqrt{2}} \ket{01}_{SF} - \ket{10}_{SF} = \ket{\psi^{-}}_{SF}.
\label{Eq.WignerDescriptionEvolutionInitialTwisted}
\end{align}
Upon measuring the same PVM as before, $\Pi_{W}=\{\ketbra{\phi^{\pm}}{\phi^{\pm}}_{SF}, \ketbra{\psi^{\pm}}{\psi^{\pm}}_{SF} \}$, now the Born's rule says his probability assignment is given by: 
\begin{align}
    \mathbb{P}_{W}(\mathcal{M}) =(0,0,0,1),
    \label{Eq.ClassicalSolutionWignersAssignmentInitialTwisted}
\end{align}
where, 
\begin{align}
    \mbox{Prob}( \phi_{+} \,\, | \,\, \ketbra{\psi^{-}}{\psi^{-}}_{SF})&=0 \nonumber , \\
    \mbox{Prob}( \phi_{-} \,\, | \,\, \ketbra{\psi^{-}}{\psi^{-}}_{SF})&=0  ,\\
    \mbox{Prob}( \psi_{+} \,\, | \,\, \ketbra{\psi^{-}}{\psi^{-}}_{SF})&=0 \nonumber  ,\\
    \mbox{Prob}( \psi_{-} \,\, | \,\, \ketbra{\psi^{-}}{\psi^{-}}_{SF})&=1. \nonumber
\end{align}
In this case, the Friend's assignment remains the same. In other words,
\begin{align}
    \mathbb{P}_{F}(\mathcal{M}) =\left(\frac{1}{2},\frac{1}{2},0,0\right),
    \label{Eq.ClassicalSolutionFriendsAssignmentInitialTwisted}
\end{align}
and $\mathcal{M}$ represents the measurement determined by the PVM $\Pi_{W}$.

Now, because $\mbox{supp}(\mathbb{P}_{W}(\mathcal{M})) \cap \mbox{supp}(\mathbb{P}_{F}(\mathcal{M})) = \emptyset$, their classical assignments are \emph{in}compatible. Similar reasoning shows that regarding Wigner and the Friend as quantum agents does not settle the matter; their (quantum) assignments would also be \emph{in}compatible. The fact that Wigner and his Friend share the same description of the initial state is fundamental, reinforcing that not agreeing on it would imply in incompatibility. 

\subsubsection{Disagreement on the inner workings of the Friend's laboratory - $NOT$ gate}

Now consider the case where Wigner and his Friend disagree on the internal procedure used to prepare the quantum state inside the laboratory. Suppose that Wigner is misinformed (or lacks precise knowledge) about the experimental setup, and believes that, instead of a Hadamard gate applied to the system, the state preparation uses an \( X \) gate. Moreover, he assumes that the subsequent interaction with the Friend is governed by a $CNOT$ gate, controlled on the system as usual:
\begin{align}
    \ket{00}_{SF} &\xmapsto{X \otimes \mathbb{1}} \ket{10}_{SF} \xmapsto{CNOT} \ket{11}_{SF}.
\label{Eq.WignerDescriptionEvolutionTwistedEvolution}
\end{align}
A hasty analysis might suggest that we are essentially dealing with the same case as before, leading to the conclusion that Wigner's and the Friend's descriptions are incompatible. However, this is incorrect. Unlike the previous scenario where Wigner predicts the completely orthogonal state $\vert{}\psi^-\rangle_{SF}$, here Wigner predicts the pure state $\vert{}11\rangle_{SF}$. When expressed in the Bell basis, Wigner's state exhibits coherences, represented by a matrix with $1/2$ and $-1/2$ entries in the upper-left $2 \times 2$ block. Crucially, the quantum support of Wigner's state is $\operatorname{span}\{\vert{}11\rangle\}$, which has a non-trivial intersection with the Friend's support, $\operatorname{span}\{\vert{}00\rangle, \vert{}11\rangle\}$. Therefore, according to Theorem 3, their descriptions are mathematically compatible, highlighting how mixing classical outcomes and distinct quantum bases can easily obfuscate the actual intersection of the agents' supports.

\subsubsection{Disagreement on the inner workings of the Friend's laboratory - Time-sensitive Compatibility}

Now consider a more subtle case in which Wigner and his Friend disagree not due to operational imprecision, but because Wigner adopts an alternative theoretical model for the state preparation. Suppose he believes that, instead of a discrete Hadamard operation, the system undergoes a continuous-time unitary evolution governed by a local Hamiltonian:  $ \mathcal{U} = e^{-i \, \frac{t}{\hbar} H} \otimes \mathbb{1}$, with Hamiltonian given by $H = \frac{\hbar \omega}{2} \sigma_y$. The initial state $\ket{00}_{SF}$ evolves in time as:
\begin{align}
   \ket{\psi(t)} = \cos\frac{\omega t}{2}\ket{00}_{SF} + \sin \frac{\omega t}{2} \ket{10}_{SF}.
    \label{Eq.WignerDescriptionEvolutionUnitary-sigmay}
\end{align}
A direct calculation shows that
\begin{align}
    \mbox{Prob}( \phi_{+} \,\, | \,\, \ketbra{\psi(t)}{\psi(t)}_{SF})&=\frac{1}{4}(1+\cos \omega t) \nonumber , \\
    \mbox{Prob}( \phi_{-} \,\, | \,\, \ketbra{\psi(t)}{\psi(t)}_{SF})&=\frac{1}{4}(1+\cos \omega t),  \\
    \mbox{Prob}( \psi_{+} \,\, | \,\, \ketbra{\psi(t)}{\psi(t)}_{SF})&=\frac{1}{4}(1-\cos \omega t) \nonumber  ,\\
    \mbox{Prob}( \psi_{-} \,\, | \,\, \ketbra{\psi(t)}{\psi(t)}_{SF})&=\frac{1}{4}(1-\cos \omega t). \nonumber
\end{align}
In this hypothetical situation, compatibility is time-sensitive. On the one hand, when $t=0$, Wigner's and his Friend's descriptions are compatible, not least because the intersection between supports is not empty, but because they coincide: 
\begin{equation}
   \mathbb{P}_{W}(\mathcal{M}))\vert_{t=0} = \mathbb{P}_{F}(\mathcal{M}). 
\end{equation}
On the other hand, when $t= \pi/\omega$, one finds 
\begin{equation}
    \mbox{supp}(\mathbb{P}_{W}(\mathcal{M}))\vert_{t=\frac{\pi}{\omega}} \cap \mbox{supp}(\mathbb{P}_{F}(\mathcal{M})) = \emptyset.
\end{equation}
Granted, as we remarked before, this is a hypothetical scenario, one in which Wigner's knowledge about the inner workings of Friend's laboratory is farfetched, but it depicts a situation in which the mere time evolution takes the descriptions from compatible to \textit{in}compatible.  

\subsubsection{Disagreement on the inner workings of the Friend's laboratory - Phase-sensitive Compatibility}

Finally, consider the case where Wigner does not reject the general structure of the Friend's experimental procedure, but suspects that an additional operation was applied, one that introduces a controlled relative phase after entangling the system and the Friend’s memory. Specifically, Wigner believes that the system undergoes a unitary transformation given by \( U(\varphi) \in \mathbb{C}^{4 \times 4} \), defined as:
\begin{equation}
U(\varphi) = \text{CP}(\varphi) \cdot \text{CNOT} \cdot (H \otimes \mathds{1}),
\end{equation}
where \( H \) is the Hadamard gate, \( \text{CP}(\varphi) \) is the Controlled Phase Gate, a two-qubit quantum gate that applies a phase shift of \( \varphi \) only when both qubits are in the \( \ket{1} \) state.\footnote{In the computational basis \( \{ \ket{00}, \ket{01}, \ket{10}, \ket{11} \} \), the gate is represented by the matrix $\mbox{diag} (1,1,1,e^{i\varphi}).$} With the identity $\mathds{1}$ acting on the second qubit—interpreted as the Friend’s memory, if one prefers that view. 

When applied to the initial state \( \ket{00}_{SF} \), this results in a relative phase between branches of the superposition, a subtle modification introduced by Wigner to account for potential coherence manipulations he suspects the Friend may have implemented. Applying \( U(\varphi) \) to the initial state \( \ket{00}_{SF} \), the system evolves as:
\begin{equation}
 U(\varphi) \ket{00}_{SF} = \frac{1}{\sqrt{2}} \left( \ket{00}_{SF} + e^{i\varphi} \ket{11}_{SF} \right) = \ket{\phi(\varphi)}_{SF}.
\end{equation}
Equivalently, this process can be expressed as: 
\begin{align}
     \nonumber \ket{00}_{SF} &\xmapsto{{H} \otimes \mathds{1}} \frac{1}{\sqrt{2}} (\ket{0}_S + \ket{1}_S) \otimes \ket{0}_F ,\nonumber\\&\xmapsto{\text{CNOT}} \frac{1}{\sqrt{2}} \left( \ket{00}_{SF} + \ket{11}_{SF} \right) ,\\&\xmapsto{CP(\varphi)} \frac{1}{\sqrt{2}} \left( \ket{00}_{SF} + e^{i\varphi} \ket{11}_{SF} \right) =\ket{\phi(\varphi)}_{SF}. \nonumber
\end{align}
In this scenario, the parameter \( \varphi \) encodes a relative phase between the computational basis states \( \ket{00}_{SF} \) and \( \ket{11}_{SF} \). 

To clarify the choice of the symbol $\phi$ for this state: when \( \varphi = 0 \), the state coincides with \( \ket{\phi^+}_{SF} \); when \( \varphi = \pi \), it becomes \( \ket{\phi^-}_{SF} \) as we see below. The state \( \ket{\phi(\varphi)}_{SF} \) remains entangled for all values of \( \varphi \), but the phase shift can lead to differing predictions for Wigner. The question is whether this phase shift can lead Wigner and his Friend to disagree about the internal dynamics of the laboratory and, potentially, result in incompatible descriptions of the same physical situation.

When \( \varphi = 0 \), the phase gate acts trivially, and the resulting state becomes:
\begin{equation}
\ket{\phi(\varphi=0)}_{SF} = \ket{\phi^+}_{SF} = \frac{1}{\sqrt{2}}(\ket{00}_{SF}+\ket{11}_{SF}).
\end{equation}
Likewise, when \( \varphi = \pi \),
\begin{equation}
\ket{\phi(\varphi=\pi)}_{SF} = \ket{\phi^-}_{SF} = \frac{1}{\sqrt{2}}(\ket{00}_{SF} - \ket{11}_{SF}).
\end{equation}
More generally, for Wigner, the pure state \( \ket{\phi(\varphi)}_{SF} \) gives rise to a density matrix \( \rho_{SF} \) featuring off-diagonal coherence terms that depend explicitly on the phase \( \varphi \). The corresponding density operator is:
\begin{align}
\nonumber \rho_{SF} &= \ketbra{\phi(\varphi)}{\phi(\varphi)}_{SF}, \\
 \nonumber    &= \frac{1}{2}(\ket{00}_{SF}+ e^{i\varphi} \ket{11}_{SF})(\bra{00}_{SF} + e^{-i\varphi} \bra{11}_{SF}), \\
     &= \frac{1}{2}(\ketbra{00}{00}_{SF}+ e^{-i\varphi} \ketbra{00}{11}_{SF} \nonumber\\& \quad+ e^{i\varphi} \ketbra{11}{00}_{SF} + \ketbra{11}{11}_{SF}).
\end{align}
The probability of getting any of the classical answers is given by the Born rule. So, now we explicitly compute the probability associated with the outcome \( \phi^+ \). Using the definition of \( \rho_{SF} \) and the Born rule, we find:
\begin{align}
    \mathds{P}_{\phi^+} &= \frac{1}{2} \Big[ |\braket{\phi^+}{00}|^2  \nonumber + e^{-i\varphi} \braket{\phi^+}{00}\braket{11}{\phi^+}  \nonumber \\
    &\quad + e^{i\varphi} \braket{\phi^+}{11}\braket{00}{\phi^+}  \nonumber + |\braket{\phi^+}{11}|^2 \Big], \nonumber \\
    &=  \frac{1}{2} \left[ \frac{1}{2} + e^{-i\varphi}  \frac{1}{2} + e^{i\varphi}  \frac{1}{2} + \frac{1}{2} \right], \nonumber  \\
    &= \frac{1+\cos\varphi}{2}.
\end{align}
These results show how the interference term governed by \( \varphi \) shifts the probability to \( \phi^+ \), highlighting the phase sensitivity of the measurement. In other words, if \( \varphi = 0 \), then \( \mathds{P}_{\phi^+} = 1 \); now, if \( \varphi = \pi \), then \( \mathds{P}_{\phi^+} = 0 \).

Similarly, we calculate the probability for obtaining the outcome \( \phi^- \) by following the same procedure:
\begin{align}
     \nonumber \mathds{P}_{\phi^-} &= \frac{1}{2} \Big[ |\braket{\phi^-}{00}|^2 
    + e^{-i\varphi} \braket{\phi^-}{00}\braket{11}{\phi^-} 
    \\& \quad+ e^{i\varphi} \braket{\phi^-}{11}\braket{00}{\phi^-} + |\braket{\phi^-}{11}|^2 \Big]]
     \nonumber,  \\ \nonumber &=  \frac{1}{2} \Big[ 1 -\frac{e^{-i\varphi}+e^{i\varphi}}{2} \Big]], \\
   &= \frac{1 - \cos\varphi}{2}. 
\end{align}
Let us now analyze how the value of \( \varphi \) affects the probability \( \mathds{P}_{\phi^-} \).
If \( \varphi = 0 \), then \( \mathds{P}_{\phi^-} = 0 \); and if \( \varphi = \pi \), then \( \mathds{P}_{\phi^-} = 1 \).

Now we turn to the Bell state \( \psi^+ \). Doing the inner product evaluations and applying the Born rule, we immediately see that:
\begin{align}
    \nonumber \mathds{P}_{\psi^+} &= \frac{1}{2} \Big[ |\braket{\psi^+}{00}|^2 
    + e^{-i\varphi} \braket{\psi^+}{00}\braket{11}{\psi^+} 
    \\&\quad + e^{i\varphi} \braket{\psi^+}{11}\braket{00}{\psi^+} 
    + |\braket{\psi^+}{11}|^2 \Big],\\
     \nonumber &= 0.
\end{align}
As expected, since neither \( \ket{00}_{SF} \) nor \( \ket{11}_{SF} \) has support on the subspace spanned by \( \ket{\psi^+} \), the corresponding probability vanishes.
The same reasoning applies to the outcome \( \psi^- \), for which we also find, that this probability is zero for all values of \( \varphi \), since there is no overlap with the support of \( \rho \). Consequently, for all values of \( \varphi \), \( \mathds{P}_{\psi^+} = 0 \) and \( \mathds{P}_{\psi^-} = 0 \).

Therefore, we have shown that:
\begin{align}
\text{Prob}(\phi^+ \mid \ket{\phi(\varphi)}\bra{\phi(\varphi)}_{SF}) &= \frac{1}{2}(1 + \cos \varphi), \\
\text{Prob}(\phi^- \mid \ket{\phi(\varphi)}\bra{\phi(\varphi)}_{SF}) &= \frac{1}{2}(1 - \cos \varphi), \\
\text{Prob}(\psi^+ \mid \ket{\phi(\varphi)}\bra{\phi(\varphi)}_{SF}) &= 0, \\
\text{Prob}(\psi^- \mid \ket{\phi(\varphi)}\bra{\phi(\varphi)}_{SF}) &= 0.
\end{align}
In this case, the Friend’s assignment remains the same. In other words,
\begin{equation}
\mathds{P}_F(\mathcal{M}) = \left( \frac{1}{2}, \frac{1}{2}, 0, 0 \right),
\end{equation}
Similarly to the time-continuous case, in this hypothetical situation, our probability distribution depends on \( \varphi \). However, in this scenario, differently from before, no choice of \( \varphi\) undermines the compatibility between Wigner and the Friend. On the one hand, when \( \varphi = 0 \), Wigner has the assignment \(\mathds{P}_W(\mathcal{M}) = (1, 0, 0, 0).\) On the other hand, when \( \varphi = \pi \), Wigner has the assignment
\(\mathds{P}_W(\mathcal{M}) = (0, 1, 0, 0).\) For any other value of \( \varphi \), the first two terms of the distribution oscillate inversely. In any case, Wigner’s and his Friend’s descriptions are always compatible, since their distributions always have a non-trivial intersection of supports.
\begin{equation}
\operatorname{supp}(\mathds{P}_W(\mathcal{M})) \cap \operatorname{supp}(\mathds{P}_F(\mathcal{M})) \neq \varnothing \,\forall\, \varphi.
\end{equation}
This result raises an intriguing question: let us consider two Wigners, hereafter $\text{Wigner}_L$ and $\text{Wigner}_R$, both positioned outside the laboratory, respectively to its left and right. Both observe the same laboratory but do not see each other. The first modulates $\varphi = 0$, while the second chooses $\varphi = \pi$. Although their descriptions are compatible with the Friend, they are not compatible with each other,
\begin{align}
    \mathds{P}_{\text{Wigner}_L}(\mathcal{M}) &= (1,0,0,0), \\
    \mathds{P}_{\text{Wigner}_R}(\mathcal{M}) &= (0,1,0,0).
\end{align} 
This result raises a crucial conceptual point regarding the nature of compatibility in this framework. Curiously, the definition of compatibility given by the authors of [23] is not transitive. That is, the fact that ``$Wigner_L$'s description is compatible with the Friend's description'' and ``the Friend's description is compatible with $Wigner_R$'s description'' does not mathematically imply that ``$Wigner_L$ is compatible with $Wigner_R$''. This lack of transitivity highlights a subtle but profound difference between our operational definition of compatibility and standard mathematical equivalence relations. We emphasize this feature because it is central to the paper's core message: within a Bayesian inference approach, compatibility is an agent-centric, pairwise alignment of beliefs rather than a universal, objective truth. It is precisely this non-transitive nature that allows multiple divergent superobservers to peacefully coexist with the local observer without triggering a genuine paradox.

\subsubsection{Wigner does not know quantum theory}

Among many other sources of disagreement, consider the exotic case in which Wigner is completely oblivious to quantum theory and understands only its classical counterpart. He knows the Friend is conducting an experiment in her laboratory; it involves some uncertainty, and he knows there are two different (classical) possible answers to it. The Friend, on the other hand, does not know that Wigner does not know quantum theory. In this extreme case, the agents do not even use the same object to describe what is happening. Wigner may be tempted to be reasonable and say that there is an even chance for each of the Friend's answers, say $\mathbb{P}_{W}=(1/2,1/2)$. The Friend, ignoring Wigner's obliviousness, assigns $\mathbb{P}_{F}=(1/2,1/2,0,0)$ to the same situation. We might be tempted to compare both assignments by appending zeroes, for example, but they are two different objects in different spaces and escape the hypothesis of theorems~\ref{Thm.ClassicalCompatibilityObjective}-\ref{Thm.QuantumCompatibility}. Not surprisingly, their assignments cannot be considered compatible—at least not in the sense of this work.
%
%
\section{RECOVERING AGREEMENT IN EUGENE WIGNER'S THOUGHT EXPERIMENT} \label{Sec.ouragreements}

Recall that, in the context of this work, agreement should be thought of as a kind of epistemic compromise among the parties (Defs.~\ref{Def.ThickNotionAgreementClassical} and~\ref{Def.ThickNotionAgreementQuantum}). How this compromise is indeed obtained is a completely different story. It is far richer and, in certain circumstances, includes the impossibility of agreeing to disagree~\cite{Aumann76}. Even if we exclude those scenarios, reaching an agreement remains complex. In the previous Sections, for example, we explored two distinct forms of compatibility depending on the shared world-view of both agents, either objective or subjective, and this is already enough to guide us to two different paths for the possibility of agreement. This Section shows what the agreed-upon probabilities would look like.  

\subsection{Subjectivist future agreement}\label{oursubjectiveagreement}

To begin with, note that theorem \ref{Thm.ClassicalCompatibilitySubjective} (and its quantum analogue) establishes that subjectivist compatibility is equivalent to the non-trivial intersection of the appropriate probability assignments' supports. Since we have already seen in the Sections \ref{SubSec.OurSolutionClassical} and \ref{SubSec.OurSolutionQuantum} that the intersection of the supports is non-trivial, compatibility is guaranteed. Therefore, we can use the steps of the sufficiency part of the theorem retroactively: starting from the non-trivial intersection of the support, we can construct the likelihood function that represents the reconciliation experiment for the original paradox.

\subsubsection{Subjetivist classical case}\label{ClassicalSubjAgree}

The proof of the theorem~\ref{Thm.ClassicalCompatibilitySubjective}'s sufficiency provides the intuition on how to construct a hypothetical experiment that compatibilises the probabilistic description of the two agents~\cite{LS14}. We choose an element from the common support, $y = \phi^+$, and define a likelihood function $\mathbb{P}(X|Y)$ for a test outcome $X \in \{0, 1\}$, where $X$ represents the classical bit,
\begin{itemize}
    \item $\mathbb{P}(X=0 | Y=\phi^+) = 1$; \quad 
    \item $\mathbb{P}(X=1 | Y=\phi^+) = 0$;
    \item $\mathbb{P}(X=0 | Y \neq \phi^+) = 0$; 
    \item $\mathbb{P}(X=1 | Y \neq \phi^+) = 1$. 
\end{itemize}
If both agreed with this likelihood function, we assume that the experiment has been performed and that the classical bit result is $X=0$. We will use Bayes' theorem to update their distributions in this way:
\begin{align}
\label{veroclass}
    \mathbb{P}_J( Y' = y' | X = 0 ) &=\frac{\mathbb{P}(X=0|Y' = y')\mathbb{P}_J(Y' = y')}{\sum_{\tilde y} \mathbb{P}(X=0|\tilde Y = \tilde y)\mathbb{P}_J(\tilde{Y} = \tilde y)} \nonumber\\&= \delta_{y',y=\phi^+}.
\end{align} We know that both agree they do not need to consider the calculations for $\mathbb{P}_W(\psi^\pm|X)$, since the result will be $0$ for both.

For Wigner, eq. \eqref{veroclass} provides
\begin{align}
    \mathbb{P}_W(\phi^+|X=0) &= \frac{\mathbb{P}(X=0|\phi^+)\mathbb{P}_W(\phi^+)}{\sum_{y} \mathbb{P}(X=0|Y = y) \mathbb{P}_W(Y = y)} \nonumber\\&= \frac{1 \cdot 1}{1} = 1. \\
    \mathbb{P}_W(\phi^-|X=0) &= \frac{\mathbb{P}(X=0|\phi^-)\mathbb{P}_W(\phi^-)}{\sum_{y} \mathbb{P}(X=0|Y = y) \mathbb{P}_W(Y = y)} \nonumber \\&= \frac{0 \cdot 0}{1} = 0.
\end{align}
Thus, Wigner's updated distribution remains the same, that is, $(1, 0, 0, 0)$.

The Friend also needs to update his knowledge via Bayesian conditioning. By applying the same procedure, we calculate the posterior probabilities for the outcome $X=0$.
\begin{align}
    \mathbb{P}_F(\phi^+|X=0) &=  \frac{1 \cdot 1/2}{1/2} = 1,  \\
    \mathbb{P}_F(\phi^-|X=0) &=  \frac{0 \cdot 1/2}{1/2} = 0.
\end{align}
The Friend's final distribution is also $(1, 0, 0, 0)$. Both agents reach \textit{agreement}, although only one of them, the Friend, has revised her description.

\subsubsection{A small detour: Cromwell's Rule}
\label{cromwell}

Now, we consider the alternative outcome: the case where $X=1$. The observers would also update their results. We analyse how Wigner deals with the fact that the classical pointer showed $1$. The calculation will not be performed explicitly because it involves an indetermination. A substantial difference is noted in Wigner's denominator calculation for this case,
\begin{align}
    \mathbb{P}(X=1) &= \sum_{y'} \mathbb{P}(X=1|Y' = y') \mathbb{P}_W(Y' = y') \nonumber \\
    &= \mathbb{P} (X=1|\phi^+)\mathbb{P}_W(\phi^+) \nonumber \\&\quad+ \mathbb{P}(X=1|\phi^-)\mathbb{P}_W(\phi^-)
    \nonumber \\&= 0.
\end{align}
The denominator is zero. For Wigner, the result $X=1$ is impossible, and the update is not defined.

Finally, we shall see what the Friend would do,
\begin{align}
    \mathbb{P}_F(\phi^+|X=1) &= \frac{\mathbb{P}(X=1|\phi^+)\mathbb{P}_F(\phi^+)}{\sum_{y} \mathbb{P}(X=1|Y = y) \mathbb{P}_F(Y = y)}  \nonumber
    \\&= \frac{0 \cdot 1/2}{(0 \cdot 1/2) + (1 \cdot 1/2)} = \frac{0 \cdot 1/2}{1/2} = 0, \\
    \mathbb{P}_F(\phi^-|X=1) &= \frac{\mathbb{P}(X=1|\phi^-)\mathbb{P}_F(\phi^-)}{\sum_{y} \mathbb{P}(X=1|Y = y) \mathbb{P}_F(Y = y)}  \nonumber\\&= \frac{1 \cdot 1/2}{(0 \cdot 1/2) + (1 \cdot 1/2)} = \frac{1 \cdot 1/2}{1/2} = 1.
\end{align}
The Friend's final distribution becomes $(0, 1, 0, 0)$. In this case, obviously, there is no agreement. However, the subjectivist compatibility theorem only requires the \textit{possibility} of agreement, which is guaranteed by the existence of the $X=0$ outcome.

The likelihood function allows them to agree on at least one classical answer, as it should. Although we have indeed reached intersubjective agreement for at least one result, we see that Wigner is ``stubborn'' in a certain sense: he cannot believe in anything that deviates from his initial distribution, whereas the Friend, in this case, is more easily convinced post-measurement. If we perceive that the problem is not merely mathematical but also represents an epistemological issue comparing unitary evolution and projection, it is strange that one side always bows to the other and the reverse never occurs. This sounds as if there were a privileged measurement interpretation over the other.

We emphasise that this is exclusively due to the Bayesian nature of our description, as evidenced by the Cromwell's rule \cite{Lindley1985,Jaynes2003}:
\emph{a Bayesian agent can never be convinced by evidence of something they initially considered absolutely impossible. Formally, if $\mathbb{P}(H) = 0$, then for any evidence $E$,
\begin{align}
    \mathbb{P}(H \mid E) = 0.
\end{align}
Therefore, the assignment of zero probabilities must be done with extreme caution.}

A careful look might suggest that the likelihood function could be biasing the result in Wigner's favor, as it uses $\phi^+$ as a parameter. This concern is valid, but in fact, it is difficult to find reasons not to adopt this approach, given that the proof of the sufficiency \cite{LS14} of Theorem \ref{Thm.ClassicalCompatibilitySubjective} (subjectivist compatibility) leads us to use a state taken from the non-trivial intersection of the support and, for this specific case, this is the only available element; therefore, the choice is, in a way, mandatory.

Are we left to settle for this agreement biased by Wigner's ``stubbornness'', or could something be done to ensure greater impartiality in the results? Before that, however, we can show the reader how to perform this construction and verification for the quantum case.

\subsubsection{Subjectivist's quantum case}\label{QuantumSubjAgree}

As we did in the last Section, in this part we will show what the quantumly agreed-upon state is. Recall that in this case, the compatibility was verified by analysing the supports of either agent's quantum assignments.  Wigner's state is a pure projector, so its support is the space spanned by its only eigenvector with a non-zero eigenvalue,
\begin{equation}
    \text{supp}(\sigma_W) = \text{span}\{\ket{\phi^+}\}.
\end{equation}
On the other hand, the state $\sigma_F$ can be rewritten in the Bell basis as
\begin{equation}
    \sigma_F = \frac{1}{2}(\ket{\phi^+}\bra{\phi^+} + \ket{\phi^-}\bra{\phi^-}).
\end{equation}
Thus, the eigenvectors with non-zero eigenvalues are $\ket{\phi^+}$ and $\ket{\phi^-}$. Therefore, its support is the subspace spanned by these vectors
\begin{equation}
    \text{supp}(\sigma_F) = \text{span}\{\ket{\phi^+}, \ket{\phi^-}\}.
\end{equation}
Again, in this case, it is easy to note that the intersection of the two supports is non trivial.
So we can rely on a construction similar to its classical counterpart and find the likelihood operator. Intuition suggests using a pure state of the common support to construct the hypothetical experiment. The only pure state in the common support is $\ket{\phi^+}$. The procedure will be as follows. We consider $X$ as a classical binary variable and define the hybrid operator as
\begin{equation}
\label{likelihoodfunctionquantum}
    \rho_{X|S} = \ketbra{0}{0}_X \otimes \ketbra{\Psi}{\Psi}_S + \ketbra{1}{1}_X \otimes (\mathbb{1}_S - \ketbra{\Psi}{\Psi}_S).
\end{equation}
If the Friend and Wigner agree on the use of this operator and observe the outcome $X=0$, they will update their states to
\begin{equation}  
\label{quantum.atualization}
\rho_{S|X=0}^{(F)} = \frac{(\rho_{X=0|S} \star \sigma_S^{(F)})}{\operatorname{Tr}_S(\rho_{X=0|S}\sigma_S^{(F)})} = \ketbra{\Psi}{\Psi}_S,
\end{equation} 
which employs the star product ($M \star N \equiv N^{1/2} M N^{1/2}$) \cite{LS14}.
Equation (\ref{likelihoodfunctionquantum}) applied here will be our recipe for finding the likelihood operator $\rho_{X|S}$ that represents an experiment asking, ``is the system in the state $\ket{\phi^+}$?'', that is, 
\begin{align}
    \rho_{X|S} &= \ket{0}\bra{0}_X \otimes \ket{\phi^+}\bra{\phi^+}_S \nonumber \\& \quad+ \ket{1}\bra{1}_X \otimes (\mathbb{1}_S - \ket{\phi^+}\bra{\phi^+}_S).
\end{align}
The operator corresponding to the outcome $X=0$ (success) is $\rho_{X=0|S} = \ket{\phi^+}\bra{\phi^+}$.

We assume that the hypothetical experiment results in $X=0$. The agents, consequently, update their states using the quantum Bayesian update rule, as explicitly stated in equation (\ref{quantum.atualization}). 

We shall see how Wigner handles this result. Wigner's initial state was $\sigma_W = \ket{\phi^+}\bra{\phi^+}$. So, we have,
\begin{align}
    \rho_{S|X=0}^{(W)} &= \frac{\rho_{X=0|S} \star \sigma_W}{\text{Tr}(\rho_{X=0|S} \sigma_W)}=  \ket{\phi^+}\bra{\phi^+}.
\end{align}
Since $\sigma_W$ is a projector, we have $(\sigma_W)^{1/2} = \sigma_W$, and the final state remains $\ket{\phi^+}\bra{\phi^+}$.

We now analyze whether the Friend agrees with this calculation. The Friend's initial state is $\sigma_F = \frac{1}{2}(\ket{\phi^+}\bra{\phi^+} + \ket{\phi^-}\bra{\phi^-})$, hence,
\begin{align}
    \rho_{S|X=0}^{(F)} &= \frac{\rho_{X=0|S} \star \sigma_F}{\text{Tr}(\rho_{X=0|S} \sigma_F)} \\
     &= \frac{(\sigma_F)^{1/2} (\rho_{X=0|S}) (\sigma_F)^{1/2}}{\text{Tr}\left(\ket{\phi^+}\bra{\phi^+} \cdot \frac{1}{2}(\ket{\phi^+}\bra{\phi^+} + \ket{\phi^-}\bra{\phi^-})\right)} \nonumber\\
    &= \frac{\frac{1}{2} \ket{\phi^+}\bra{\phi^+}}{1/2} = \ket{\phi^+}\bra{\phi^+}.
\end{align}
Both agents reach an agreement, updating their beliefs to one and the same final state: $\ket{\phi^+}\bra{\phi^+}$, thus resolving the apparent contradiction.

We now turn to analyse what would have happened had the outcome of the hypothetical experiment been $X=1$. This outcome corresponds to the likelihood operator
\begin{equation}
    \rho_{X=1|S} = \mathbb{1}_S - \ket{\phi^+}\bra{\phi^+}.
\end{equation}
For Wigner, his initial belief is $\sigma_W = \ket{\phi^+}\bra{\phi^+}$. To apply the update rule, we first calculate the probability of Wigner obtaining the result $X=1$. This probability is the denominator of Bayes' formula,
\begin{align}
    \mathbb{P}(X=1) &= \text{Tr}(\rho_{X=1|S} \sigma_W) \nonumber \\
    &= \text{Tr}\left( (\mathbb{1}_S - \ket{\phi^+}\bra{\phi^+}) \cdot \ket{\phi^+}\bra{\phi^+} \right) \nonumber \\ 
    &= \text{Tr}\left( \ket{\phi^+}\bra{\phi^+} - \ket{\phi^+}\bra{\phi^+} \right) = 0. 
\end{align}
Similarly to the classical case, the probability of Wigner obtaining this result is zero. From his point of view, this event is impossible. Since one cannot condition on a zero-probability event, the Bayesian update rule is undefined for Wigner. For the Friend, there is a $1/2$ probability of obtaining the $X=1$ outcome, and the update is well-defined. In fact, 
\begin{align}
    \rho_{S|X=1}^{(F)} &= \frac{(\sigma_F)^{1/2} (\rho_{X=1|S}) (\sigma_F)^{1/2}}{\text{Tr}(\rho_{X=1|S} \sigma_F)} \nonumber \\
    &=  \frac{\frac{1}{2} \ket{\phi^-}\bra{\phi^-}}{\frac{1}{2} \text{Tr}(\ket{\phi^-}\bra{\phi^-})} = \ket{\phi^-}\bra{\phi^-}. 
\end{align}
The Friend updates his belief from a mixed state to $\ket{\phi^-}$ (or $\ket{\phi^-}\bra{\phi^-}$). All in all, when $X=1$, the agents do not reach an agreement. 

We stress that this does not invalidate the concept of compatibility. Compatibility requires only the existence of \textit{at least one} outcome that leads to agreement. Since the outcome $X=0$ satisfies this condition, the initial states are considered compatible. The purpose of the subjectivist test is to demonstrate that reconciliation is possible in principle, not that it is guaranteed for all outcomes of all possible experiments.

\subsection{Interlude: Benefit of the Doubt}\label{benefitofdoubt}

We have seen that the result holds for both the classical and quantum cases. Although we have a scenario where reconciliation—and, therefore, agreement—is possible, Wigner's inflexibility in accepting a scenario where he is wrong in his prediction indicates that his prior may have been stubbornly constructed and does little to help us make an unbiased decision in this discussion. We now have sufficient reason to believe that Wigner should be more willing to grant at least a grain of benefit of the doubt before sitting down to talk with his Friend. Any sensible rational observer would do so.

Thus, let us consider that Wigner supposes his state is not so pure; that, instead of an idealized unitary, his initial state is prepared via a noisy channel that inserts an infinitesimally small parameter $\varepsilon$, representing the benefit of the doubt for the possibility of the post-measurement state being $\ket{\phi^-}$. Wigner's certainty is still mostly placed on the classical outcome signalling $\ket{\phi^+}$, with a chance $(1-\varepsilon)$, where $\varepsilon \ll 1$; that is, nearly 100\%. Nevertheless, for the sake of a neutral and impartial discussion, he considers that the preparation channel might have an infinitesimal noise level, or even that it is indeed a unitary in an expanded space (in a Stinespring representation) that accounts for some interaction with the environment. 

In this modified scenario, for Wigner, the initial state of the system and the Friend is represented by the density matrix $\rho_{\text{initial}} = \ket{00}\bra{00}_{SF}$. However, instead of an ideal unitary evolution (the composition of Hadamard and CNOT gates), we consider a noisy physical process, $\mathcal{E}$. This channel transforms the initial pure state into a mixed state, which represents Wigner's belief, now including the \textit{benefit of the doubt} $\varepsilon$,
\begin{equation} \label{eq:real_wigner_alt}
\rho_{\text{initial}} \xrightarrow{\mathcal{E}_{\text{noisy}}} \rho'_{W},
\end{equation}
or, more explicitly,
\begin{equation} \label{eq:real_wigner}
\ket{00}\bra{00}_{SF}\overset{\mathcal{E}_{\text{noisy}}}{\longmapsto} (1-\varepsilon)\ket{\phi^+}\bra{\phi^+} + \varepsilon\ket{\phi^-}\bra{\phi^-}.
\end{equation}
In practice, firstly we start in the state $\ket{00}_{SF}$, which evolves to $\ket{\phi^+}$ according to the steps in  \eqref{Eq.WignerDescriptionEvolution} --- $H \otimes \mathbb{1}$ followed by a $CNOT$. Then, $\ket{\phi^+}\bra{\phi^+}$ is submitted to the channel $\mathcal{E}_{\text{noisy}}$. In the computational basis, $\mathcal{E}_{\text{noisy}}$ is represented by the Kraus operators $\kappa_0 = \sqrt{1-\varepsilon}\mathbb{I}\otimes \mathbb{I}$, $\kappa_1 = \sqrt{\varepsilon}\sigma_3 \otimes \mathbb{I}$. 

The probability distribution in the classical sense representing Wigner's perspective is $\mathbb{P}'_{W} = (1-\varepsilon, \varepsilon, 0 ,0)$ and, therefore, the Friend and Wigner do not merely have a non-trivial intersection of supports, but rather both supports coincide. This leads us to have not just one, but two distinct elements (or, in the quantum case, two pure states) to construct our experiment (the likelihood function), and both are capable of leading the two observers to agreement on a classical answer. 

There is something far more interesting that we can extract from assignments that allow for the benefit of the doubt. However, to discuss it, we must first take an interlude and examine how \emph{state improvement} works in an inferential scenario.

\subsection{Improvement}\label{improvement}

We call the update of a state assignment in light of another agent's state assignment a \textit{state improvement}~\cite{Herbut2004, LS14}. In this Section, we will explore classical and quantum assignments in a Wigner's friend scenario with a twist given by the benefit of the doubt.  

\subsubsection{Improvement of classical distributions}\label{classsicaldistributionsimprovement}

Classically, suppose Freddy (a decision-maker and a proxy for the Friend) assigns a prior state $\mathbb{P}_{0}(Y)$ to the variable of interest $Y$. Lacking specialised knowledge about $Y$, his prior state might be something simple, such as a uniform distribution over $Y$'s outputs. To improve this assignment, Freddy consults someone he assumes to be an expert: Wanda, a proxy for Wigner. She provides him with an opinion in the form of a new state $\mathbb{P}_{1}(Y)$. 

Since Freddy has access to neither the data nor the rationale used by Wanda, he can only rely on her provided summary, that is, on $\mathbb{P}_{1}(Y)$. To perform a Bayesian update, Freddy must treat Wanda's assignment as data, thus constructing a likelihood function $\mathbb{P}_{0}(R \mid Y)$, where $R$ is a random variable encompassing all possible reports Wanda could provide. In practice, $R$ is usually restricted to well-parameterised families of states (for instance, Gaussian states or a discrete set of options). Furthermore, Freddy can take into account factors such as Wanda's reliability or her accuracy in previous predictions. 

In this scenario, the likelihood function is no longer an idealised experiment for testing hypotheses, but a model of trust or testimony (an epistemological assessment). In words, the definition is: ``Given the hypothesis that the true state of nature is $Y$, what is the probability that the Agent (the 'Expert') will report to me that their belief is $R$?''. To a certain extent, the nature of this function is, in a way, even more introspective, as it depends on how reliable, accurate, or biased an agent believes the other agent also ought to be.

With this in mind, this is how Freddy can update his prior state using Bayes' theorem, obtaining~\cite{LS14},
\begin{equation}
\mathbb{P}_{0}(Y \mid R = \mathbb{P}_{1}) = \frac{\mathbb{P}_{0}(R = \mathbb{P}_{1} \mid Y)\,\mathbb{P}_{0}(Y)}{\mathbb{P}_{0}(R = \mathbb{P}_{1})},
\label{aprimo}
\end{equation}
where the denominator is given by
\begin{equation}
\mathbb{P}_{0}(R = \mathbb{P}_{1}) = \sum_{Y} \mathbb{P}_{0}(R = \mathbb{P}_{1} \mid Y)\,\mathbb{P}_{0}(Y).
\label{denon}
\end{equation}
Another factor to mention is that this improvement is relative: the observer who improves their distribution must infer that the other observer is an expert. In a case where both start from the same prior, this is easier to discern: an expert might be the one who performed more measurements and gathered more data about the system. By improving their distribution, the less-informed agent should end up with the same distribution as the expert, without needing to worry about which experiments were performed or which measurements were taken. In a case where both start from different \textit{priors}, the agent who improves their assignment is not obliged to agree entirely with the expert. They might merely expand their knowledge with new information from the expert, yet continue to have a divergent distribution—likely compatible but not in total agreement. Obviously, this is a two-way street: both can investigate what they would learn if they took the other as the expert.

It is time to return to Wigner's thought experiment. Let \(W\) (Wigner) and \(F\) (the Friend) be two observers who assign probability distributions over the same space of classical outcomes $\mbox{Out}(Y)=\{y_1 = \phi^+, y_2= \phi^-, y_3=\psi^+, y_4=\psi^-\},$ which correspond to the four outcomes of a Bell test. We denote by \(\mathbb{P}_W\) and \(\mathbb{P}_F\) the distributions that each observer assigns to \(Y\).

\vspace{12pt}
{\centering
\textit{1.a) The Friend is an Expert for Stubborn Wigner ($W \leftarrow F$)} \par}
\vspace{12pt}

We shall perform the improvement of classical distributions considering a stubborn Wigner. Let us see what happens when Wigner, with $\mathbb{P}_W(Y) = (1,\, 0,\, 0,\, 0)$ as a prior, performs the improvement taking the Friend as an expert with $\mathbb{P}_F(Y) = (1/2, 1/2, 0, 0)$. The probability, according to Wigner's belief, that $\mathbb{P}_{F}$ reports $(1/2,1/2,0,0)$ conditioned on $Y$ is,
\begin{align}
\mathbb{P}_W(R = \mathbb{P}_F \mid Y) 
    &= \left( \tfrac{1}{2}, \tfrac{1}{2}, 0, 0 \right).
\end{align}
Consequently, we have,
\begin{align} 
\mathbb{P}_W(Y|R=\mathbb{P}_F) &= \frac{(\frac{1}{2} , \frac{1}{2} , 0, 0) \odot (1, 0, 0, 0)}{\sum_{Y'} \mathbb{P}_W(R=\mathbb{P}_F|Y') \cdot \mathbb{P}_W(Y')} 
\nonumber\\ &= \frac{(\frac{1}{2} , 0 , 0, 0)}{\frac{1}{2} \cdot 1 +\frac{1}{2} \cdot 0 +0 \cdot 0 + 0 \cdot 0}\nonumber
\\ &= (1, 0, 0, 0), 
\end{align}
where $\odot$ is the \textit{Hadamard} entrywise product \footnote{The Hadamard product ($\odot$) is the element-by-element operation. For two distributions $\bm{F}$ and $\bm{W}$ in the same distribution space, the resulting vector $\bm{I}$ is defined by $\odot: \bm{F} \times \bm{W} \to \bm{I}$, such that $I(y) \coloneqq F(Y=y) \cdot W(Y=y)$ for each component $y$.}. 

Once again, we witness the materialization of Wigner's obstinacy. Ultimately, Wigner does not change his mind, even knowing that his Friend is an expert on the subject. The Friend's data serves only for him to reaffirm what he already believed.

\vspace{12pt}
{\centering
\textit{1.b) Stubborn Wigner is an Expert for the Friend ($F \leftarrow W$)} \par}
\vspace{12pt}

The question is: does the same occur when the Friend performs an improvement when they consider (the stubborn) Wigner as an expert?
Explicitly we obtain,
\begin{align}
\mathbb{P}_F(\phi^+ \mid R=\mathbb{P}_W)  &= \frac{\mathbb{P}_W(\phi^+) \cdot \mathbb{P}_F(\phi^+)}{\sum_{Y'} \mathbb{P}_W(Y') \mathbb{P}_F(Y')} 
= \frac{1 \cdot \tfrac{1}{2}}{1 \cdot \tfrac{1}{2}} = 1, 
\end{align}but since $\mathbb{P}_W(\phi^-)= \mathbb{P}_W(\psi^+)= \mathbb{P}_W(\psi^-)=0$ then $\mathbb{P}_F(Y \mid R=\mathbb{P}_W) =0$ for all $Y \neq \phi^+$.
Therefore, the Friend's updated belief, after incorporating the expert's assignment, becomes
\begin{align}
\mathbb{P}_F(Y \mid R= \mathbb{P}_W) = (1,\, 0,\, 0,\, 0).
\end{align}
If the Friend maintains an open mind, opening the laboratory and sharing information leads her to agree with Wigner. But what happens if we relax Wigner's absolute certainty by introducing the uncertainty parameter $\varepsilon$? In other words, what changes when we insert the benefit of the doubt?

\vspace{12pt}
{\centering
\textit{1.c) The Friend is an expert for the Open-Minded Wigner ($W \leftarrow F$)} \par}
\vspace{12pt}

In the stubbornness situation, Wigner assigned probability one to a single Bell state. Now, we relax this assumption,
\begin{align}
    \mathbb{P}'_W(Y) = (1 - \varepsilon,\, \varepsilon,\, 0,\, 0),
    \quad \text{with} \quad 0 < \varepsilon \ll 1,
\end{align}
to represent Wigner's more open-minded distribution. We then perform the improvement taking the Friend as an expert. Our likelihood function will be $\mathbb{P}_W(R=\mathbb{P}_F\mid Y) = (\varepsilon, 1 - \varepsilon, 0, 0)$. This is because there is not a very wide variety of options to test here. It makes no sense to contradict the zeros of $\psi^+$ or $\psi^-$, and the two distributions that assume certainty about $\phi^+$ and $\phi^-$ are biased and have already been studied. Although this occurred in another context, here they would provide us with a similar result, so they are not of interest to us. It makes much more sense to test one of the two families of distributions with $\varepsilon$ and its complement (in the sense of summing to $1$) $1-\varepsilon$. We are assuming the one that opposes Wigner's prior and will shortly present the reason for this choice.

The Bayesian improvement rule, with Wigner taking the Friend as an expert, reads
\begin{align}
    \mathbb{P}'_W(Y \mid R = \mathbb{P}_F)
    =
    \frac{\mathbb{P}'_W(R = \mathbb{P}_F \mid Y) \odot \mathbb{P}'_W(Y)}
    {\sum_{Y'} \mathbb{P}'_W(R=\mathbb{P}_F \mid Y') \odot \mathbb{P}'_W(Y')}.
\end{align}
With more details,
\begin{align} 
 \mathbb{P}'_W(Y \mid R=\mathbb{P}_F) &= \frac{(\varepsilon, 1 - \varepsilon, 0, 0) \odot (1 - \varepsilon, \varepsilon , 0 , 0 )}{\sum_Y (\varepsilon, 1 - \varepsilon, 0, 0) \odot (1 - \varepsilon, \varepsilon , 0 , 0 )} 
 \nonumber\\
  &= \frac{(\varepsilon (1 - \varepsilon),  (1 - \varepsilon)  \varepsilon , 0, 0)}{2 \varepsilon  ( 1-\varepsilon)} 
 \nonumber\\
  &= (1/2, 1/2, 0, 0) \nonumber \\
  &= \mathbb{P}_F(Y).
\end{align} 
That is, we finally have our first scenario in which Wigner agrees with his Friend's prior after opening the laboratory and sharing data. Therefore, this likelihood function reveals a singular behavior: the more certain Wigner is, the more extreme the likelihood function must be so that the improvement allows Wigner to be convinced by his Friend. This is the motivation behind choosing this function.

\vspace{12pt}
{\centering
\textit{1.d) Open-Minded Wigner is an Expert for the Friend ($F \leftarrow W$)} \par}
\vspace{12pt}

Let us now observe what happens if we maintain Wigner's (un)certainty relaxation parameter for the reverse improvement. We calculate the Friend's updated belief state upon receiving information from Wigner.

The Friend constructs a model $\mathbb{P}_F(R = \mathbb{P}'_W | Y)$ of how Wigner would arrive at his report. The assumption is that Wigner's report faithfully reflects the external reality he observes, with his small open-mindedness. A possible model is based on the following rationale: $(i)$ If the reality is $Y=\phi^+$, Wigner's report will be exactly $\mathbb{P}'_W$; and $(ii)$ if the reality is $Y=\phi^-$, Wigner would report a different distribution (for example: $(\varepsilon, 1-\varepsilon, 0, 0)$); therefore, the probability of him having made the report $\mathbb{P}'_W$ is zero. Formally, the likelihood function is,
\begin{align}
    \mathbb{P}_F(\text{Wigner reports } \mathbb{P}'_W | Y=\phi^+) &= 1, \\
    \mathbb{P}_F(\text{Wigner reports } \mathbb{P}'_W | Y=\phi^-) &= 0.
\end{align}
With the Bayes' rule,
\begin{equation}
    \mathbb{P}_F(Y | R=\mathbb{P}'_W) = \frac{\mathbb{P}_F(R=\mathbb{P}'_W | Y) \cdot \mathbb{P}_F(Y)}{\mathbb{P}_F(R=\mathbb{P}'_W)}.
\end{equation}
Finally, the Friend can perform the calculation of the \textit{posterior},
\begin{align}
    \mathbb{P}_F( Y |R=\mathbb{P}'_W) &= \frac{\mathbb{P}_F(R=\mathbb{P}'_W|Y)\mathbb{P}_F(Y)}{\sum_{y'} \mathbb{P}_F(R=\mathbb{P}'_W|Y' = y') \mathbb{P}_F(Y' = y')}.
 \end{align}
More explicitly,
 \begin{align}
    \mathbb{P}_F(\phi^+|R=\mathbb{P}'_W) &= \frac{\mathbb{P}_F(R=\mathbb{P}'_W|\phi^+)  \mathbb{P}_F(\phi^+)}{\sum_{y'} \mathbb{P}_F(R=\mathbb{P}'_W|Y' = y') \mathbb{P}_F(Y' = y')} 
    \nonumber\\&=
    \frac{1 \cdot 1/2}{1 \cdot 1/2 + 0 \cdot 1/2} = 1; 
    \\
    \mathbb{P}_F(\phi^-|R=\mathbb{P}'_W) &= \frac{\mathbb{P}_F(R=\mathbb{P}'_W|\phi^-)  \mathbb{P}_F(\phi^-)}{\sum_{y'} \mathbb{P}_F(R=\mathbb{P}'_W|Y' = y') \mathbb{P}_F(Y' = y')} ,
    \nonumber\\&=
    \frac{0 \cdot 1/2}{1 \cdot 1/2 +0 \cdot 1/2} = 0. 
\end{align}The Friend's improved belief state, after considering Wigner's report, is,
\begin{equation}
    \mathbb{P}_F(Y |  R= \mathbb{P}'_W) = (1, 0, 0, 0).
\end{equation}
The information provided by Wigner, even if not of absolute certainty, is decisive. This strong asymmetric evidence resolves the Friend's initial uncertainty, leading him to adopt Wigner's main standpoint. 

\subsubsection{Improvement for Quantum States}
\label{quantumstatesimprovement}

Improving quantum assignments resembles its classical counterpart~\cite{LS14}. The major difference is that we work with hybrid conditional states and the quantum Bayes' theorem~\cite{LS13}. 

In our scenario, Freddy is now interested in a quantum region $S$, to which he assigns a prior state $\rho^{(0)}_S$, and Wanda announces her expert state assignment $\sigma^{(1)}_S$. Analogously to the classical case, Freddy will treat Wanda's announcement as new data and construct a classical random variable $R$ that takes Wanda's possible state assignments as values. Constructing a sample space for all possible states is, again, technically subtle, but in practice, attention can be restricted to well-parameterised families. Freddy's likelihood is now a hybrid conditional state $\rho^{(0)}_{R|S}$, and he updates his prior state assignment via the hybrid Bayes' theorem to obtain:
\begin{equation}
\rho^{(0)}_{S|R=\sigma^{(1)}_S} = \rho^{(0)}_{R=\sigma^{(1)}_S |S} \star \rho^{(0)}_S \left( \rho^{(0)}_{R=\sigma^{(1)}_S} \right)^{-1},
\end{equation}
where $\rho^{(0)}_{R=\sigma^{(1)}_S} = \text{Tr}_S \left( \rho^{(0)}_{R=\sigma^{(1)}_S |S} \rho^{(0)}_S \right)$. 

\textbf{Remark:} Note that the same methodology can be applied when Freddy consults more than one expert: Wanda, Theo, etc. Freddy must only construct a likelihood function $\mathbb{P}(R_1, R_2, \dots |Y )$ in the classical case or a likelihood operator $\rho_{R_1 R_2 \dots|B}$ in the quantum case, where $R_1$ represents Wanda's state assignment, $R_2$ represents Theo's state assignment, and so on. He then applies the appropriate version of Bayes' theorem to condition on the state assignments announced by the experts. For simplicity, we will only focus on the case with two agents.

We will start by assuming that Wigner's certainty is absolute, that is, $\varepsilon = 0$. In this case, we have the following initial beliefs (prior): Wigner with state $\sigma_W = \ket{\phi^+}\bra{\phi^+}$, and the Friend with state $\sigma_F = \frac{1}{2}(\ket{\phi^+}\bra{\phi^+} + \ket{\phi^-}\bra{\phi^-})$. Below, we will analyse either case of quantum improvement: (1) the Friend is an expert for Wigner, and (2) Wigner is an expert for the Friend.

\vspace{12pt}
{\centering
\textit{2.a) The Friend is an Expert for Stubborn Wigner ($W \leftarrow F$)} \par} 
\vspace{12pt}

Here, Wigner acts as the decision-maker and the Friend as the expert. The received report is $R = \sigma_F$. Wigner constructs a likelihood operator $\rho_{R|S}$ that models his trust in the Friend. Given that Wigner possesses absolute certainty that the state is $\ket{\phi^+}$, he must assign zero probability to any event that contradicts this belief. The calculation via the quantum Bayes' rule is given by,
\begin{equation}
    \rho_{S|R=\sigma_F}^{(W)} = \frac{\rho_{R=\sigma_F|S} \star \sigma_W}{\text{Tr}(\rho_{R=\sigma_F|S} \sigma_W)}.
\end{equation}
Since $\sigma_W$ is a pure projector, the star product in the numerator collapses into the state itself scaled by a scalar factor (the matrix element of the likelihood),
\begin{align}
\rho_{S|R=\sigma_F}^{(W)} 
&= \frac{\sigma_{W}^{\frac{1}{2}} \rho_{R=\sigma_F|S} \sigma_{W}^{\frac{1}{2}}}{\operatorname{Tr}(\rho_{R=\sigma_F|S} \ket{\phi^+}\bra{\phi^+})} \nonumber \\
&= \frac{\sqrt{\ket{\phi^+}\bra{\phi^+}} \rho_{R=\sigma_F|S} \sqrt{\ket{\phi^+}\bra{\phi^+}}}{\bra{\phi^+}\rho_{R=\sigma_F|S} \ket{\phi^+}} 
\nonumber\\
&= \frac{\ket{\phi^+}\bra{\phi^+} \rho_{R=\sigma_F|S} \ket{\phi^+}\bra{\phi^+}}{\bra{\phi^+} \rho_{R=\sigma_F|S} \ket{\phi^+}}
\nonumber\\
&= \ket{\phi^+}\bra{\phi^+}. 
\end{align}
As we discussed earlier, Wigner's stubbornness shields him from his friend's beliefs. He improves based on the report, but since the report is orthogonal to his initial belief, it remains unchanged.

\vspace{12pt}
{\centering
\textit{2.b) The Friend takes a stubborn Wigner as an Expert ($F \leftarrow W$)} \par}
\vspace{12pt}

In this case, the Friend is the decision-maker and Wigner is the expert. The report is $R = \sigma_W = \ket{\phi^+}\bra{\phi^+}$. The Friend constructs a likelihood model based on the premise that Wigner, as a superobserver, possesses the correct theory of unitary evolution. In other words, $(i)$ if the state is $\ket{\phi^+}$, Wigner will report $\sigma_W$ with probability 1; and $(ii)$ if the state is $\ket{\phi^-}$, Wigner would never report $\sigma_W$ (probability $0$). This defines the following likelihood operator,
\begin{align}
\rho_{R=\sigma_W|B} = \ket{\phi^+}\bra{\phi^+}.
\end{align}
Consequently,
\begin{align}
    \rho_{S|R=\sigma_W}^{(F)} &= \frac{\rho_{R=\sigma_W|S} \star \sigma_F}{\text{Tr}(\rho_{R=\sigma_W|S} \sigma_F)} \nonumber \\
    &= \frac{\sigma_F^{1/2} \rho_{R=\sigma_W|B} \sigma_F^{1/2}}{\operatorname{Tr}\left(\ket{\phi^+}\bra{\phi^+} \cdot \frac{1}{2}(\ket{\phi^+}\bra{\phi^+} + \ket{\phi^-}\bra{\phi^-})\right)} \nonumber \\
    &= \frac{\frac{1}{2} \ket{\phi^+}\bra{\phi^+}}{1/2} = \ket{\phi^+}\bra{\phi^+}. 
\end{align}
The Friend was, then, convinced by Wigner's certainty. The improvement process changes the Friend's initial statistical mixture into the pure state proposed by Wigner. Curiously, the likelihood as $\phi^-$ would drive the Friend's improvement toward a pure state with no agreement and no compatibility with Wigner's state—a total disconnection of beliefs.

\vspace{12pt}
{\centering
\textit{2.c) The Friend is an expert for Open-Minded Wigner ($W \leftarrow F$)}\par}
\vspace{12pt}

Now Wigner no longer assigns a rank-one projector and begins to consider a statistical mixture for his assignment, albeit a highly biased one, allowing for a small probability that the system is in the orthogonal state $\ket{\phi^-}$. Wigner's prior state becomes:
\begin{equation}
\sigma_W^{(\varepsilon)} = (1-\varepsilon)\ket{\phi^+}\bra{\phi^+} + \varepsilon\ket{\phi^-}\bra{\phi^-},
\end{equation}
with $0 < \varepsilon \ll 1$. The Friend's state remains the same.

Wigner receives the report $R=\sigma_F$. For the improvement to work in this scenario of mitigated stubbornness, the likelihood function must follow the same logic found in the classical case. That is, the probability of the Friend's beliefs must compensate for Wigner's own initial belief. We construct the likelihood operator analogous to the classical vector $(\varepsilon, 1-\varepsilon, 0, 0)$:
\begin{align}
    \rho_{R=\sigma_F|S} = \varepsilon \ket{\phi^+}\bra{\phi^+} + (1-\varepsilon) \ket{\phi^-}\bra{\phi^-}.
    \label{Eq.QuantumWignerOpenMindAss}
\end{align}
Since all operators are diagonal in the Bell basis $\{\ket{\phi^{\pm}}, \ket{\psi^{\pm}}\}$, they commute. Thus, the star product reduces to the usual matrix product. So, applying the quantum Bayes' theorem, one finds
\begin{align}
\rho_{S|R=\sigma_F}^{(W)} &= \frac{(\sigma_W^{(\varepsilon)})^{1/2} \rho_{R=\sigma_F|S} (\sigma_W^{(\varepsilon)})^{1/2}}{\text{Tr}\left( (\sigma_W^{(\varepsilon)})^{1/2} \rho_{R=\sigma_F|S} (\sigma_W^{(\varepsilon)})^{1/2} \right)} \nonumber \\
&= \frac{\varepsilon(1-\varepsilon)\ket{\phi^+}\bra{\phi^+} + \varepsilon(1-\varepsilon)\ket{\phi^-}\bra{\phi^-}}{\varepsilon(1-\varepsilon) \left[ \text{Tr}(\ket{\phi^+}\bra{\phi^+}) + \text{Tr}(\ket{\phi^-}\bra{\phi^-}) \right]} \nonumber \\
&= \frac{1}{2}(\ket{\phi^+}\bra{\phi^+} + \ket{\phi^-}\bra{\phi^-})  = \sigma_F. 
\end{align}
Therefore, the improved state is $\sigma_F$. As we discussed in the classical case, the introduction of the benefit of the doubt ($\varepsilon > 0$), combined with an appropriately constructed likelihood, allows Wigner to ``move away'' from his quasi-certainty and agree with the Friend's statistical mixture.

\vspace{12pt}
{\centering
\textit{2.d) Open-Minded Wigner is an Expert for the Friend ($F \leftarrow W$)} \par}
\vspace{12pt}

Keeping Wigner's open-minded prior as in eq.~\eqref{Eq.QuantumWignerOpenMindAss}, we will study how the Friend updates her state $\sigma_F$ on his report. The likelihood assumption here is that Wigner's report is a strong indicator that his preferred state ($\ket{\phi^+}$) is the true one. If the state were $\ket{\phi^-}$, Wigner (being a reliable observer) would be unlikely to report a distribution so heavily weighted on $\ket{\phi^+}$. We model this situation by assuming that the report $\sigma_W^{(\varepsilon)}$ flags almost exclusively $\ket{\phi^+}$. The simplified likelihood operator is the projector:
\begin{equation}
    \rho_{R=\sigma_W^{(\varepsilon)}|S} = \ket{\phi^+}\bra{\phi^+}.
\end{equation}
The update calculation follows,
\begin{align}
\rho_{S|R=\sigma_{W}^{(\epsilon)}}^{(F)}&=\frac{\rho_{R=\sigma_{W}|S}*\sigma_{F}}{Tr(\rho_{R=\sigma_{W}|S}\sigma_{F})}\nonumber\\
&=\frac{\sigma_{F}^{1/2}|\phi^{+}\rangle\langle\phi^{+}|\sigma_{F}^{1/2}}{Tr(|\phi^{+}\rangle\langle\phi^{+}|\cdot\frac{1}{2}(|\phi^{+}\rangle\langle\phi^{+}|+|\phi^{-}\rangle\langle\phi^{-}|))}\nonumber\\
&=\frac{[\frac{1}{\sqrt{2}}\Pi_{\phi^{\pm}}]|\phi^{+}\rangle\langle\phi^{+}|[\frac{1}{\sqrt{2}}\Pi_{\phi^{\pm}}]}{\frac{1}{2}Tr(|\phi^{+}\rangle\langle\phi^{+}|)}\nonumber\\
&=|\phi^{+}\rangle\langle\phi^{+}|,
\end{align}
where $\Pi_{\phi^{\pm}} = \vert{}\phi^{+}\rangle\langle\phi^{+}\vert{} + \vert{}\phi^{-}\rangle\langle\phi^{-}\vert{}$ denotes the identity operator restricted to the two-dimensional subspace spanned by $\{\vert{}\phi^{+}\rangle, \vert{}\phi^{-}\rangle\}$, acting here as a projector.

Again, we observe that Wigner's evidence is strong enough to drive the Friend's original beliefs, leading her to agree with Wigner's pure state assignment, $|\phi^{+}\rangle$.

Again, we observe that Wigner's evidence is strong enough to drive the Friend's original beliefs, leading her to agree with Wigner's pure state assignment, $\ket{\phi^+}$. 

All in all, under the condition of non-zero support (guaranteed by $\varepsilon$), it is possible to reconcile the agents' perspectives, whether by the Friend's ``surrender'' to Wigner's world-view or by the Friend's conviction when Wigner keeps an open mind. In the Appendix \ref{ourpooling}, we will explore another possible reconciling technique. 

\section{Conclusions}\label{Sec.Conclusions}

In Eugene Wigner's Original Thought Experiment, the central point of contention is the vague nature of the triad agreement-compatibility-irreconcilability. The paradox usually arises from the fact that Wigner's and his Friend's descriptions are different, thus incompatible, therefore irreconcilable---and consequently paradoxical. In this work, we took a Bayesian detour and provided a clearer definition of classical and quantum compatibility. This definition was aimed at resolving the blurred lines surrounding the triad, and by separating them, we concluded that there is no paradox in Eugene Wigner's Original Thought Experiment.

By making use of our Bayesian definition (basically the BFM criterion/condition), we reconcile Wigner's and the Friend's discrepant descriptions in several ways. Within the subjectivist framework, we created a classical experiment in which, when used by Wigner and his Friend, both agree on at least one classical result; we also proposed a quantum experiment, represented by a likelihood operator, through which they agree on the classical answer that Wigner does not consider absurd. Still within the subjectivist standpoint, and bringing to the table the concept of state improvement, we analysed the various means for this reconciliation: in the classical case, we explored Wigner consulting his Friend and the Friend consulting Wigner, followed by the insertion of the benefit of the doubt, where we repeated the entire process. Finally, we proceeded analogously in the quantum case: Wigner to the Friend, the Friend to Wigner, and both considering the benefit of the doubt. We also present a notion within the objectivist framework via a classical joint distribution and its analogous hybrid joint state---see Appendices \ref{jointdistribution} and \ref{hybridstate}--- as well as three distinct ways of performing state pooling---see Appendix \ref{ourpooling}.

These conclusions come with a cost. We had to reframe Wigner's Original Formulation into a problem of multi-agent Bayesian inference. Wigner wants to describe what is happening to an agent he has full control of, but not information about; and likewise, one may want to ask the Friend what she thinks Wigner's description is. Each agent is reasoning and inferring about situations in which they lack full information. The spirit of our reframing builds upon the working-idea that quantum theory is a particular theory of probabilistic assignments. One in which the agents' bets are density operators, PVMs and CPTP maps—a scholarship initiated in refs.~\cite{Leifer14,LS13,LS14}. A sounder notion of compatibility and the possibility to analyse agreement are direct by-products of that scholarship. 

However, the ideas promoted by this scholarship introduce their own set of controversial points. It is important to note that compatibility, as defined here, is not quantifiable. We are unable to quantify the degree of incompatibility between the two assignments. They either have trivial support or they do not. One may want to expand this concept by examining, say, the inner product of the two distributions, which would likely provide a smoother definition. Whether this new quantifier truly captures compatibility in the Bayesian sense will certainly be explored elsewhere. Note, also, that there is a crucial assumption in all the definitions of compatibility, especially in its objective sense. It is assumed that there exists another random variable that can be jointly measured with the original ones and such that the given marginals are the correct ones; a question that strongly resembles contextuality and that will also be adequately addressed in future work. We cannot avoid mentioning that the definition of compatibility we employ here is fragile enough so that small variations in the Wigner scenario would lead to irreconcilable differences between his and the Friend's descriptions---as we make clear in the main text. 

Recall that we have also offered critiques of the aforementioned compatibility criterion: the dependence of the agreement on the initial state of the quantum system and on how the dynamics unfold; the temporal sensitivity (perceived in the continuous evolution over time); and the lack of transitivity. We accept that such factors make the criterion less robust, and it may be possible to robustify the definition considering operational symmetries. For instance, one may consider the permutation orbit of a given distribution and compare equivalence classes via this orbit rather than individual probability vectors. But at least as far as we are concerned, all such variations stem from a lack of agreement between the agents, which simultaneously says that there is no paradoxical situation and that agreement plays a fundamental role in inference problems—be they classical or quantum~\cite{LD22}. Finally, one may ask whether we can deploy a similar rationale to the many extensions of Wigner's Original Thought Experiment. Because they are not directly formulated as a disagreement or a discrepancy between the friends' probabilistic descriptions, we are unsure whether we can apply the reasoning we advance here.   

\begin{acknowledgments}

The authors would like to thank Olival Freire Jr. for his careful review of the manuscript and for suggesting appropriate references. CD thanks the hospitality of the Institute for Quantum Studies at Chapman University and the camaraderie he has found at the Fundação Maurício Grabois. CD also thanks Raphael Drumond for his input on this work. This work was supported by CNPq through a grant from the Programa Conhecimento Brasil
 (Linha 1 and Linha 2). CD was supported by Grant 63209 from the John Templeton Foundation. The opinions expressed in this publication are those of the authors and do not necessarily reflect the views of the John Templeton Foundation. Julio Cesar would like to express his gratitude to the Universidade Federal de Juiz de Fora and to the Brazilian funding agency, CAPES, for the support provided. And to the master's program in physics at the Programa de Pós-Graduação - Física, Department of Physics, UFJF. JC also thanks Vinícius Valle for his comments on the work. The authors dedicate this paper to the memory of José Roberto Júnior.    
\end{acknowledgments}

%

\appendix
\section{Quantum Compatibility: objective and subjective}\label{SubApp.DefQuantumCompatibility}
\label{quantumdef}

\begin{definition}[Quantum Objective Bayesian Compatibility]
Two quantum states $\sigma_{S}^{W}$ and $\sigma_{S}^{F}$ over a quantum region $S$ are \emph{compatible} whenever it is possible to find a pair of random variables $X_1$ and $X_2$ and a hybrid state $\rho_{SX_1X_2}$, such that $\sigma_{S}^{W}$ can be obtained by Bayesian conditioning on $X_1 = x_1$ for some value $x_1 \in \mbox{Out}(X_1)$, and $\sigma_{S}^{F}$ can be obtained by Bayesian conditioning on $X_2 = x_2$ for some value $x_2 \in \mbox{Out}(X_2)$. In other words,
\begin{enumerate}
    \item [(a)] $\sigma_{S}^{W} = \rho_{S | X_1 = x_1}.$
    \item [(b)] $\sigma_{S}^{F} = \rho_{S | X_2= x_2}.$
    \item [(c)] $\rho_{X_1=x_1 | X_2 = x_2} \neq 0$
\end{enumerate}
For a pair of outcomes $(x_1,x_2) \in \mbox{Out}(X_1) \times \mbox{Out}(X_2)$.
    \label{Def.QuantumObjectiveCompatibility}
\end{definition} 
\textbf{Remark:} We demand that $\rho_{X_1=x_1 | X_2 = x_2} \neq 0$ so that there is a possibility for both outcomes to be obtained simultaneously, which greatly constrains the set of random variables where we can pick $X_1$ and $X_2$ from—if they are meant to be of any meaning in quantum theory.

\begin{definition}[Quantum Subjective Bayesian Compatibility]
Two quantum states $\sigma_{S}^{W}$ and $\sigma_{S}^{F}$ over a quantum region $S$ are \emph{compatible} whenever it is possible to find a random variable $X$ and a conditional state $\rho_{X \vert S}$ such that there exists a particular value $x' \in \mbox{Out}(X)$ such that,
\begin{enumerate}
    \item [(a)] $\mbox{Tr}_{S} \left( \rho_{X=x' \vert S} \sigma_{S}^{J}  \right) \neq 0$.
    \item [(b)] $\rho_{S | X =x'}^{W} = \rho_{S | X =x'}^{F}$
\end{enumerate}
The conditional hybrid state in item (b) is defined, for $J \in \{F,W\}$ by the quantum Bayes's theorem,
\begin{align}
    \rho_{S | X = x}^{J} := \left( \rho_{X = x \vert S} \, \star \, \sigma_{S}^{J} \right)\mbox{Tr}_{S}\left(\rho_{X = x \vert S } \, \star \, \sigma_{S}^{J} \right)^{-1}. 
\end{align}
    \label{Def.QuantumSubjectiveCompatibility}
\end{definition} 

\section{Construction of the joint distribution of classical variables for classical objectivist compatibility (agreement in a virtual past)}\label{jointdistribution}

The resolution of the Wigner’s Friend paradox through a Bayesian framework demonstrates that contradiction is avoided once compatibility is verified. While subjectivist agreement is achieved via a shared experiment, objectivist agreement traditionally requires a common prior from which all individual perspectives are derived. Although Wigner and the Friend lack a common prior in the original setup, the equivalence established by Theorems \ref{Thm.ClassicalCompatibilityObjective} and \ref{Thm.QuantumCompatibility} allows for a retrospective construction. If the agents' assignments have a non-trivial support intersection, they are compatible; being compatible, there must exist a joint classical-quantum state that unifies their information.

We don't see this as possible agreement in the past, but we define as agreement in a possible past, or agreement in the ``virtual past''. Even if such a state was not part of the initial narrative, mathematics permits the derivation of a ``virtual past'' state that justifies their current compatible descriptions. This approach preserves the original assignments while providing a robust objectivist grounding for their reconciliation. 

Let $\mathbb{P}(Y, F, W)$ be the joint probability we seek. As established many times in this work, the values that $Y$ can assume are equivalent to the Bell basis $ Y = \{ \phi^+, \phi^-, \psi^+, \psi^- \}.$ The Friend's marginal distribution is $\mathbb{P}_F(Y) = (1/2, 1/2, 0, 0),$ representing a 50\% chance of measuring $\phi^+$ and a 50\% chance of measuring $\phi^-$. Wigner's distribution is $\mathbb{P}_W(Y) = (1, 0, 0, 0),$ representing absolute certainty of measuring $\phi^+$. Classical compatibility has already been verified previously.

We now follow the construction of this distribution based on the sufficiency procedure of Theorem \ref{Thm.ClassicalCompatibilityObjective} (Classical Objectivist Bayesian) that one could check more thoroughly in \cite{LS14}, decomposing our distribution according to the equations below. Given that $\mathbb{P}_1(Y)$ and $\mathbb{P}_2(Y)$ have a non-trivial intersection of supports, we can find a probability distribution $\mathbb{P}_0(Y)$ such that:
\begin{equation}
\label{decomp1}
    \mathbb{P}_F(Y) = p_F \mathbb{P}_0(Y) + (1-p_F) \mathbb{P}'_F(Y),
\end{equation}
\begin{equation}
\label{decomp2}
    \mathbb{P}_W(Y) = p_W \mathbb{P}_0(Y) + (1-p_W) \mathbb{P}'_W(Y),
\end{equation}
where $0 < p_F, p_W \leq 1$, and the functions $\mathbb{P}'_F(Y)$ and $\mathbb{P}'_W(Y)$ are normalized probability distributions. We decompose the assignments into a convex combination to facilitate the construction of the statistical model. This decomposition allows us to construct the classical variables $X_1$ and $X_2$ and the joint probability $\mathbb{P}(Y, X_1, X_2)$ such that $\mathbb{P}(X_1=x_1, X_2=x_2) \neq 0$. We define $X_1$ and $X_2$ as discrete variables taking values in the set $\{0,1\}$. We establish the marginal probabilities as $\mathbb{P}(X_1=0) = p_F$ and $\mathbb{P}(X_2=0) = p_W$, so that the complementary probabilities are $\mathbb{P}(X_1=1) = 1-p_F$ and $\mathbb{P}(X_2=1) = 1-p_W$, ensuring normalization.

Now applying this to our specific case, we seek a part $\mathbb{P}_0(Y)$ common to both Wigner and the Friend,
\begin{align}
    \mathbb{P}_0(Y) = (1, 0, 0, 0) = \mathbb{P}_W(Y),
\end{align}for this specific case.

We have the general form of the decomposition as a convex combination of distributions:
\begin{align}
    \mathbb{P}_i(Y) = p_i \mathbb{P}_0(Y) + (1-p_i)\mathbb{P}'_i(Y),
\end{align}
where $0 < p_F, p_W \leq 1$, and $\mathbb{P}'_F$ and $\mathbb{P}'_W$ are, in themselves, normalized probability distributions. Since the premise is that there is a non-trivial intersection of supports, $\mathbb{P}_F(Y)$ and $\mathbb{P}_W(Y)$ always possess a part in $\mathbb{P}_0(Y)$; therefore, $p_F$ and $p_W$ are strictly positive.

Then, for Wigner, the statistical weight $p_W$ will be:
\begin{align}
    \mathbb{P}_W(Y) &= 1 \cdot \mathbb{P}_0(Y) + (1-1)\mathbb{P}'_W(Y) = \mathbb{P}_0(Y), \\&\implies p_W = 1.
\end{align}
$p_F$ can also be evaluated in the Friend's distribution,
\begin{align}
    \mathbb{P}_F(Y) &= \frac{1}{2} \mathbb{P}_0(Y) + \left(1 - \frac{1}{2}\right)\mathbb{P}'_F(Y)\Rightarrow p_F = \frac{1}{2},
\end{align}
where $\mathbb{P}'_F(Y) = (0, 1, 0, 0)$, corresponding to $\phi^-$.

We now define our conditionals' cases:
\begin{enumerate}
    \item \label{item1} $\mathbb{P}(Y | F=0, W=0) = \mathbb{P}_0(Y) = (1, 0, 0, 0)$;
    \item \label{item2} $\mathbb{P}(Y | F=0, W=1) = \mathbb{P}'_F(Y) = (0, 1, 0, 0)$.
\end{enumerate}
The other cases have weight $0$ because, if $p_W=1$, the term accompanied by $(1-p_W)=0$ vanishes and will not appear in our considerations. The case $F=1$ and $W=1$ is irrelevant for this construction:
\begin{align}
    \mathbb{P}(Y | F=1, W=0) &= \mathbb{P}'_W(Y) = \text{irrelevant}; \\
    \mathbb{P}(Y | F=1, W=1) &= N(Y) = \text{irrelevant}.
\end{align}
Irrelevant means $0$-weight in this case.

We obtained $p_F=1/2$ and $p_W=1$ such that it is possible to obtain the following probabilities for the auxiliary variables:
\begin{align}
    a) \quad & \mathbb{P}(F=0, W=0) = p_F p_W =1/2; \\
    b) \quad & \mathbb{P}(F=0, W=1) = (1-p_F) p_W=1/2; \\
    c) \quad & \mathbb{P}(F=1, W=0) = p_F (1-p_W)=0; \\
    d) \quad & \mathbb{P}(F=1, W=1) = (1-p_F)(1-p_W)=0
\end{align}
With the conditionals and the joint probabilities of the auxiliary variables, we have the elements to find the total joint distribution 
\begin{equation}\label{conj}
    \mathbb{P}(Y, F, W) = \mathbb{P}(Y | F, W) \mathbb{P}(F, W).
\end{equation}
Since $p_Fp_W > 0$, it follows that $\mathbb{P}(F,W)$ can be non-zero in the cases of interest.

Substituting the relevant terms into Eq. (\ref{conj})—that is, for $F=0, W=0$ we have conditionals' case \ref{item1}. with the probability of item $a)$, and for $F=0, W=1$ we have the conditionals' case \ref{item2}. with the probability of item $b)$, since the others vanish—we finally obtain:
\begin{equation}
    \mathbb{P}(Y, F, W) = (1, 0, 0, 0)\frac{1}{2} + (0, 1, 0, 0)\frac{1}{2}.
\end{equation}
Now, we verify this result. First, we demonstrate that we can recover the Friend's distribution, $\mathbb{P}_F(Y) = \mathbb{P}(Y|F=0)$, from this joint distribution.

The probability of the Friend obtaining result $0$ (success in the mixture) will be the denominator $\mathbb{P}(F=0)$ and the joint probability marginalized over $W$ will be our numerator so, the conditional is, therefore:
\begin{align}
\label{verified1}
    \mathbb{P}(Y | F=0) &= \frac{\mathbb{P}(Y, F=0)}{\mathbb{P}(F=0)} \nonumber \\&= \frac{\sum_W \mathbb{P}(Y | F=0, W)\mathbb{P}(F=0,W) }{\sum_W \mathbb{P}(F=0, W)}  \nonumber \\ 
    &= \frac{\mathbb{P}(Y|0,0)\mathbb{P}(F=0, W=0)}{\mathbb{P}(F=0, W=0) + \mathbb{P}(F=0, W=1)} \nonumber \\
    &\quad+\frac{\mathbb{P}(Y|0,1)\mathbb{P}(F=0, W=1)}{\mathbb{P}(F=0, W=0) + \mathbb{P}(F=0, W=1)} \nonumber \\
    &= \frac{(1, 0, 0, 0)\frac{1}{2} + (0, 1, 0, 0)\frac{1}{2}}{\frac{1}{2} + \frac{1}{2}} \nonumber \\
    &= \frac{(1/2, 1/2, 0, 0)}{1} = \mathbb{P}_F(Y). 
\end{align}
Finally, we show that it is possible, via marginalization, to recover Wigner's distribution, $\mathbb{P}_W(Y) = \mathbb{P}(Y \vert W=0)$, from the joint distribution. Here, our denominator will be the probability of Wigner obtaining result $0$ and our numerator will be the joint probability marginalized over $F$. Together with the conditional results, we obtain, 
\begin{align}
\label{verified2}
    \mathbb{P}(Y | W=0) &= \frac{\mathbb{P}(Y, W=0)}{\mathbb{P}(W=0)} \nonumber \\
    &= \frac{\sum_F \mathbb{P}(Y, F, W=0)}{\sum_F \mathbb{P}(F, W=0)} \nonumber \\
    &= \frac{\mathbb{P}(Y|F=0, W=0)\mathbb{P}(F=0, W=0)}{\mathbb{P}(F=0, W=0) + \mathbb{P}(F=1, W=0)} \nonumber \\
    &\quad+ \frac{  \mathbb{P}(Y|F=1, W=0)\mathbb{P}(F=1, W=0)}{\mathbb{P}(F=0, W=0) + \mathbb{P}(F=1, W=0)} \nonumber \\
    &= \frac{(1, 0, 0, 0)\frac{1}{2} + \mathbb{P}(Y|1,0) \cdot 0}{\frac{1}{2} + 0} \nonumber \\
    &= (1, 0, 0, 0) = \mathbb{P}_W(Y).
\end{align}
We have thus demonstrated that it is possible to recover each observer's information from the joint information representing the prior distribution of a possible common past.

\section{Construction of the joint hybrid state for quantum objectivist compatibility (agreement in a virtual past)}\label{hybridstate}

In this Section, we explicitly construct the joint hybrid state $\rho_{SFW}$ that demonstrates the compatibility between the perspectives of the Friend and Wigner. We shall use a construction similar to the classical counterpart, inspired by the sufficiency of Theorem \ref{Thm.ClassicalCompatibilitySubjective} to derive definition \ref{Def.QuantumObjectiveCompatibility} from the Appendix~\ref{SubApp.DefQuantumCompatibility}.

The Friend's perspective on system $S$ is given by the state $\sigma_{S}^{(F)} = \frac{1}{2}\ket{00}\bra{00} + \frac{1}{2}\ket{11}\bra{11}$, and Wigner's is given by the following state $\sigma_{S}^{(W)} = |\phi^{+}\rangle\langle\phi^{+}|$, where $|\phi^{+}\rangle = \frac{1}{\sqrt{2}}(\ket{00} + \ket{11})$.

The first step consists of confirming the compatibility between the assignments. Again, the compatibility between the assignments is immediately verified by noting that Wigner's pure state, $|\phi^{+}\rangle$, resides in the subspace spanned by the Friend's statistical mixture. This satisfies the necessary condition for the construction of the hybrid state.

Following a procedure analogous to the classical one we performed, we decompose the states $\sigma_{S}^{(F)}$ and $\sigma_{S}^{(W)}$ with respect to a state $\mu_S$ belonging to the common support. The natural choice for $\mu_S$ is Wigner's state itself,
 \begin{align}
    \mu_S = \sigma_{S}^{(W)} = |\phi^{+}\rangle\langle\phi^{+}|.
 \end{align}
The decomposition takes the form $\sigma_{S} = p\mu_{S} + (1-p)\eta_{S}$. For Wigner's state ($\sigma_{S}^{(W)}$), the decomposition is trivial, with weight $p_W = 1$:
 \begin{align}
    \sigma_{S}^{(W)} = 1 \cdot |\phi^{+}\rangle\langle\phi^{+}| + (1-1)\eta_{S}^{(W)} = |\phi^{+}\rangle\langle\phi^{+}|.
 \end{align}
The residual state $\eta_{S}^{(W)}$ is irrelevant, as its weight is null.

For the Friend's state ($\sigma_{S}^{(F)}$), we express their mixture as a combination of $\mu_S$ and a residual state $\eta_S^{(F)}$. Using the Bell basis, we rewrite the Friend's state as an equiprobable mixture of $|\phi^+\rangle$ and $|\phi^-\rangle = \frac{1}{\sqrt{2}}(\ket{00} - \ket{11})$,
 \begin{align}
    \sigma_S^{(F)} = \frac{1}{2} (|\phi^+\rangle\langle\phi^+| + |\phi^-\rangle\langle\phi^-|).
 \end{align}
Comparing this with the form $\sigma_{S}^{(F)} = p_F \mu_{S} + (1-p_F)\eta_{S}^{(F)}$, we directly identify the weight $p_F = \frac{1}{2}$ and the residual state as $\eta_{S}^{(F)} = |\phi^-\rangle\langle\phi^-|$, which is a valid quantum state.

We then define the hybrid conditional states for the quantum region $S$, given the results of the classical binary variables $F$ and $W$ (values $\{0, 1\}$). We associate the original states of the Friend and Wigner with the result `$0$` of their respective variables:
\begin{itemize}
    \item $\rho_{S|F=0, W=0} = \mu_S = |\phi^+\rangle\langle\phi^+|$;
    \item $\rho_{S|F=0, W=1} = \eta_S^{(F)} = |\phi^-\rangle\langle\phi^-|$;
    \item $\rho_{S|F=1, W=0} = \eta_S^{(W)}$ (irrelevant, as $\mathbb{P}(F=1, W=0)=0$);
    \item $\rho_{S|F=1, W=1} = \nu_S$ (irrelevant).
\end{itemize}
The joint classical probability distribution over $F$ and $W$ is defined by the weights $p_F = 1/2$ and $p_W = 1$:
\begin{align}
    \mathbb{P}(F=0, W=0) &= p_F p_W = 1/2;  \\
    \mathbb{P}(F=0, W=1) &= (1-p_F)p_W = 1/2; \\
    \mathbb{P}(F=1, W=0) &= p_F(1-p_W) = 0; \\
    \mathbb{P}(F=1, W=1) &= (1-p_F)(1-p_W)= 0. 
\end{align}
The corresponding classical state is $\rho_{FW} = \frac{1}{2}\ket{00}\bra{00}_{FW} + \frac{1}{2}\ket{01}\bra{01}_{FW}$.

The joint hybrid state is constructed by combining the conditional states with the joint classical probability:
 \begin{align}
 \rho_{SFW} = \sum_{f,w} &\mathbb{P}(F=f, W=w)\times \nonumber \\ &\times \rho_{S|F=f, W=w} \otimes \ket{f,w}\bra{f,w}_{FW}.
 \end{align}
Substituting the calculated values, we obtain,
\begin{align}
    \rho_{SFW} = \frac{1}{2}(|\phi^+\rangle\langle\phi^+|_{S} \otimes \ket{00}\bra{00}_{FW}) + \nonumber \\ +\frac{1}{2}(|\phi^-\rangle\langle\phi^-|_{S} \otimes \ket{01}\bra{01}_{FW}). 
\end{align}
Finally, we verify the consistency of the result by recovering the original states via conditioning. For the Friend's state ($\sigma_{S}^{(F)} = \rho_{S|F=0}$), the marginal probability of obtaining the result `$0$` is the denominator and the joint state for $S, F=0$ is the numerator. then we have the conditional state results in:
\begin{align}
\label{verified3}
    \rho_{S|F=0} &= \frac{\rho_{S, F=0}}{\mathbb{P}(F=0)} \nonumber \\
    &= \frac{\mathbb{P}(F=0, W=0)\rho_{S|F=0, W=0} }{\mathbb{P}(F=0, W=0) + \mathbb{P}(F=0, W=1)} \nonumber \\
     &\quad+ \frac{\mathbb{P}(F=0, W=1)\rho_{S|F=0, W=1}}{\mathbb{P}(F=0, W=0) + \mathbb{P}(F=0, W=1)} \nonumber \\
    &= \frac{\frac{1}{2}\ket{\phi^+}\bra{\phi^+} + \frac{1}{2}\ket{\phi^-}\bra{\phi^-}}{\frac{1}{2} + \frac{1}{2}} \nonumber \\
    &= \frac{1}{2}\left ( \ket{\phi^+}\bra{\phi^+} + \ket{\phi^-}\bra{\phi^-}\right ) = \sigma_{S}^{(F)}.
\end{align}
Analogously, to recover Wigner's state ($\sigma_{S}^{(W)} = \rho_{S|W=0}$), the marginal probability of Wigner obtaining the result $0$ is our denominator and the joint state of $S$ for $W=0$ is the numerator. At last the conditional state is given by,
\begin{align}
\label{verified4}
    \rho_{S|W=0} &= \frac{\rho_{S, W=0}}{\mathbb{P}(W=0)} \nonumber \\
    &= \frac{\mathbb{P}(F=0, W=0)\rho_{S|F=0, W=0}}{\mathbb{P}(F=0, W=0) + \mathbb{P}(F=1, W=0)} \nonumber \\
     &\quad+ \frac{\mathbb{P}(F=1, W=0)\rho_{S|F=1, W=0}}{\mathbb{P}(F=0, W=0) + \mathbb{P}(F=1, W=0)} \nonumber \\
    &= \frac{\frac{1}{2}|\phi^+\rangle\langle\phi^+| + 0 \cdot \rho_{S|F=1, W=0}}{\frac{1}{2} + 0} \nonumber \\
    &= \frac{\frac{1}{2}|\phi^+\rangle\langle\phi^+|}{\frac{1}{2}} = |\phi^+\rangle\langle\phi^+| = \sigma_{S}^{(W)}.
\end{align}
This confirms that the constructed hybrid state correctly unifies the perspectives.

\section{Pooling methods}
\label{ourpooling}

\subsection{Pooling as a agreement method}\label{pooling}

In Bayesian theory, the purpose of states is to provide information for rational decision-making, which, in turn, must be performed based on all relevant available evidence. 

The fact that another agent assigns a particular state can be relevant evidence and may cause you to alter your state assignment, as we saw in Sec.\ref{improvement}, even in the case where both state assignments are the same. For example, if both you and I assign the same high probability to some event, then telling you my state assignment may cause you to assign a higher probability if you believe that my reasons for assigning a high probability are valid and independent of yours.

It is worth noting that, if two agents indeed have different state assignments, then they may have different preferences regarding the choices available in decision-making scenarios. In practice, decisions often have to be made as a group. This motivates the need for methods to combine state assignments into a single assignment that accurately represents the beliefs, information, or knowledge of the group as a whole. This problem is called state pooling. 

In the classical case, both state improvement and pooling have been studied extensively (see \cite{GenestZidek1986} and \cite{Jacobs1995} for reviews). From the analysis of the literature, it becomes evident that there is no hope of establishing a universal pooling rule --- that is, a simple functional of the different state assignments --- that is applicable to all cases. As detailed in \cite{GenestZidek1986}, the search for such a formula faces 'impossibility theorems', where desirable consistency axioms conflict, preventing the existence of a single solution that satisfies all logical criteria. 

Furthermore, the analysis in \cite{Jacobs1995} demonstrates that fixed pooling rules (such as linear pooling) are often insufficient as they fail to capture the structure of dependence and correlation between sources, which requires a more sophisticated treatment of information. Instead of a rigid rule, we use a general methodology for combining states in our problem, taken directly from \cite{LS14}, which serves both classical and quantum cases. This methodology is grounded in the application of Bayesian conditioning (a 'Supra-Bayesian' approach originally from \cite{KeeneyRaiffa1976}), allowing individual assignments to be treated as data and explicitly modelling their correlations for state pooling.

This approach requires agents to place themselves in the position of a neutral decision-maker. Although their ability to do so is not guaranteed, doing so reduces the pooling problem to a type of state improvement, that is, the state of the neutral decision-maker is conditioned on all the state assignments of the other agents, and the result is used as the pooled state.

We take the general methodology for quantum pooling from \cite{LS14}, but it can be seen more extensively in \cite{GenestZidek1986}. As with compatibility, our approach to these problems is to draw parallels with the classical case using conditional states and to derive our results through a grounded application of Bayesian conditioning. This is not part of the main scope of this work, but still, this represents an improvement over previous approaches \cite{Spekkens07, Brun2002, Jacobs2002, Herbut2004, PoulinBlumeKohout2003, Jacobs2005}, which use principles formulated for specific cases.

\subsection{Supra-Bayesian pooling}
\label{oursuprapooling}

Here, we provide the construction of the combined state $\mathbb{P}$ via quantum state pooling. As previously discussed, Wigner and the Friend do not share a common prior. To enable the pooling of their beliefs, we introduce a fictional prior $\mathbb{P}_0(Y)$, attributed to a hypothetical neutral decision-maker, say, Debbie, where $Y$ may assume the values $\{\phi^{\pm}, \psi^{\pm}\}$.

Each agent possesses a likelihood function over $Y$,
\begin{align}
    &\mathbb{P}(X_W \mid Y) \quad \text{(for Wigner)},\\ 
    &\mathbb{P}(X_F \mid Y) \quad \text{(for the Friend)}.
\end{align}
From $\mathbb{P}_0(Y)$ and the likelihoods, we have,
\begin{align}
    \mathbb{P}_i(Y) = \mathbb{P}_0(Y \mid X_i = x_i).
\end{align}
We need to evoke a conditional independence hypothesis. That is, we assume that the reports from the Friend and Wigner are conditionally independent given $Y$, namely,
\begin{align}
    \mathbb{P}_0(x_W, x_F \mid Y) = \mathbb{P}_0(x_W \mid Y) \, \mathbb{P}_0(x_F \mid Y).
\end{align}
After all, once we know $Y$, the Friend's report gives us no new information regarding Wigner's report, and vice-versa.

We adopt the principle of indifference for the choice of the prior. Since both agree that $\psi^+$ and $\psi^-$ are impossible, we can adopt,
\begin{align}
    \mathbb{P}_0(Y) = \left(\tfrac{1}{2}, \tfrac{1}{2}, 0, 0 \right).
\end{align}
Then the likelihood function of the agents are:
\begin{align}
    \mathbb{P}(X_W \mid Y) =
    \begin{cases}
        1, & Y = \phi^+ ;\\
        0, & Y = \phi^-;
    \end{cases}\\
    \mathbb{P}(X_F \mid Y) =
    \begin{cases}
        1/2, & Y = \phi^+ ;\\
        1/2, & Y = \phi^-.
    \end{cases}
\end{align}
Now, to perform the pooling via Supra-Bayesianism, we have,
\begin{equation}
    \mathbb{P}_{\text{supra}}(Y) = \frac{\mathbb{P}_0(Y) \mathbb{P}(X_W \mid Y) \mathbb{P}(X_F \mid Y)}
    {\sum_{Y'} \mathbb{P}_0(Y') \mathbb{P}(x_W \mid Y') \mathbb{P}(x_F \mid Y')}. 
\end{equation}
With details, 
\begin{align}    
    \mathbb{P}_{\text{supra}}(\phi^+) &=
    \frac{\frac{1}{2} \cdot 1 \cdot \frac{1}{2}}{\frac{1}{2} \cdot 1 \cdot \frac{1}{2}
    + \frac{1}{2} \cdot 0 \cdot \frac{1}{2}}
    = 1,\\
    \mathbb{P}_{\text{supra}}(\phi^-) &=
    \frac{\frac{1}{2} \cdot 0 \cdot \frac{1}{2}}{\frac{1}{2} \cdot 1 \cdot \frac{1}{2}
    + \frac{1}{2} \cdot 0 \cdot \frac{1}{2}}
    = 0,
\end{align}
implying 
\begin{equation}
\mathbb{P}_{\text{supra}}(Y) = (1, 0, 0, 0).    
\end{equation}
This means that, even with a hypothetical common prior, Wigner's position dominates the pooling, eliminating the disagreement.

\subsection{Multiplicative or logarithmic pooling}
\label{ourlogpooling}

Classically, a multiplicative opinion pool (or logarithmic pool) has the following form,
\begin{align}
    \mathbb{P}_{\text{mult}}(Y) = c \prod_{i=1}^{m} \mathbb{P}_i(Y)^{w_i}.
\end{align}
Applying this to the two agents, the component-by-component calculation would be,
\begin{align}
= c \left( \left(\frac{1}{2}\right)^{w_F} \cdot 1^{w_W}, \left(\frac{1}{2}\right)^{w_F} \cdot 0^{w_W}, 0, 0 \right).
\end{align}
The normalization constant $c$ is defined as
\begin{align}
c = \frac{1}{\sum_{Y} \prod_{i=1}^{m} \mathbb{P}_i(Y)^{w_i}},
\end{align}
where the weights $w_i$ must satisfy the conditions:
\begin{align}
0 < w_i < 1 \quad \text{and} \quad \sum_{i=1}^{m} w_i = 1.
\end{align}
Multiplicative pooling generally results in a pooled state that is more ``peaked'' than that of any of the individual agents' states. Here, ``more peaked'' is used in the sense that the concentration level of a probability distribution or a quantum state around specific values is higher.

Normalizability implies that multiplicative pooling can only be applied to states that are jointly compatible. This means there exists at least one value $y$ of $Y$ such that $\mathbb{P}_J(Y=y) > 0$ for all $J$. Any such value has non-zero weight in $\mathbb{P}_{\text{mult}}(Y)$, which ensures that the pooling is compatible with each agent's individual assignment. As we will see below, multiplicative pooling is useful from a Bayesian objectivist perspective where all agents start with a shared uniform prior, and their differences result from collecting different data from the system.

For cases where the shared state is non-uniform, we have the generalized multiplicative pooling,
\begin{align}
\label{gmult}
    \mathbb{P}_{\text{gmult}}(Y) = c \prod_{i=0}^{m} \mathbb{P}_i(Y)^{w_i},
\end{align}
which includes the extra state $\mathbb{P}_0(Y)$ representing the shared prior information.

Unlike linear pools, the way to generalize a multiplicative pool to its quantum version is not immediately easy to see due to the product of states in Eq. \eqref{gmult}, which does not have a unique generalization due to non-commutativity.

\subsection{Linear pooling}
\label{ourlinearpooling}

The linear opinion pool method is a straightforward way to combine the probability distributions of different agents into a single consensual distribution. It works by creating a weighted average of individual beliefs.

The formula for the linear opinion pool is
\begin{align}
    \mathbb{P}_{\text{lin}}(Y) = \sum_{i=1}^{m} w_i \mathbb{P}_i(Y).
\end{align}
The calculation proceeds component-by-component for each outcome of $Y$:
\begin{align}
\mathbb{P}_{\text{lin}}(\phi^{+}) &= w_W \cdot \mathbb{P}_W(\phi^{+}) + w_F \cdot \mathbb{P}_F(\phi^{+}) ,\\&= w_W \cdot 1 + w_F \cdot \frac{1}{2},\\
\mathbb{P}_{\text{lin}}(\phi^{-}) &=w_W \cdot \mathbb{P}_W(\phi^{-}) + w_F \cdot \mathbb{P}_F(\phi^{-}) ,\\&= w_W \cdot 0 + w_F \cdot \frac{1}{2},\\
\mathbb{P}_{\text{lin}}(\psi^{\pm}) &= 0.
\end{align}
The final combined distribution vector is:
\begin{align}\label{linear.pooling.formula}
\mathbb{P}_{\text{lin}}(Y) = \left( w_W + \frac{w_F}{2}, \quad \frac{w_F}{2}, \quad 0, \quad 0 \right).
\end{align}
Linear pooling acts as a diplomatic solution, producing a compromise that depends on the chosen weights ($w_W + w_F = 1$).

For example, let us consider the case of equal confidence, that is, $w_W = 0.5$ and $w_F = 0.5$. Then, according to \eqref{linear.pooling.formula},
\begin{align}
\mathbb{P}_{\text{lin}}(Y) = (0.75, 0.25, 0, 0).
\end{align}
The combined belief approaches Wigner's, but retains part of the Friend's uncertainty.

Suppose now that one expects the Friend to be an expert. That way, one choose $w_W = 0.1$ and $w_F = 0.9$. Then, the pooled state \eqref{linear.pooling.formula} reads
\begin{align}
\mathbb{P}_{\text{lin}}(Y) = (0.55, 0.45, 0, 0).
\end{align}
The final belief is strongly influenced by Wigner's certainty, but a small doubt from the Friend persists.

\end{document}